\documentclass[aps,onecolumn,nopacs,nofootinbib,floatfix,superscriptaddress]{revtex4}
\usepackage{graphicx}
\usepackage{amsfonts}
\usepackage{amssymb}
\usepackage{amsbsy}
\usepackage{hyperref}
\usepackage{amsmath}
\usepackage{mathrsfs}
\usepackage{latexsym}
\usepackage{natbib}
\usepackage{bm}
\usepackage{subfigure} 
\usepackage{color}
\usepackage{wasysym}
\usepackage{mathbbol}
\usepackage{bigints}
\allowdisplaybreaks
\usepackage[normalem]{ulem}
\usepackage[dvipsnames]{xcolor}

\newcommand{\indra}[1]{\textcolor{violet}{[{\bf Indranil}: #1]}}

\def\erf{\mathcal{E}\text{rf}}

\begin{document}

\title{Entanglement harvesting for different gravitational wave burst profiles with and without memory}

\author{Subhajit Barman}
\email{subhajit.barman@physics.iitm.ac.in}
\affiliation{Centre for Strings, Gravitation and Cosmology, Department of Physics, Indian Institute of Technology Madras, Chennai 600036, India}

\author{Indranil Chakraborty}
\email{indranil.chakraborty9@gmail.com}
\affiliation{Department of Physics, Indian Institute of Technology Bombay, Mumbai 400076, India}

\author{Sajal Mukherjee}
\email{sajal.mukherjee@pilani.bits-pilani.ac.in} 
\affiliation{Department of Physics, Birla Institute of Technology and Science - Pilani, Rajasthan 333031, India}
\affiliation{Astronomical Institute of the Czech Academy of Sciences, Bo\v{c}n\'{i} II 1401/1a, CZ-141 00 Prague, Czech Republic}

\begin{abstract}
\noindent
 In the present article, we study how different gravitational wave (GW) burst profiles in linearized gravity, with and without the asymptotic memory, may influence the harvesting between two static Unruh-DeWitt detectors. To this end, we investigate the following burst profiles-- Gaussian, sech-squared, Heaviside step function, and tanh. Out of these, the first two bursts contain no memory, while the latter two consist of a non-vanishing memory effect. We find that in all of these cases, entanglement harvesting is possible, and it decreases with the increasing distance between detectors and the detector transition energy. We observe that the harvesting differs qualitatively based on the presence or absence of the memory, which is prominent in a low transition energy regime. With memory, the harvesting keeps increasing with decreasing transition energy, while without memory, it tends to reach finite values. Furthermore, for the two burst profiles without memory, longer bursts correspond to greater harvesting in the low detector transition energy regime, and this characteristic is reversed for larger transition energy. Meanwhile, for the tanh-type profile with memory, harvesting is always greater for shorter bursts. We discuss various implications of our findings.

\end{abstract}

\date{\today}

\maketitle
%%%%%%%%%%%
\section{Introduction}\label{sec:Introduction}
%%%%%%%%%%%
The arena of relativistic quantum information (RQI) \cite{Reznik:2002fz, Floreanini:2004, FuentesSchuller:2004xp, Ball:2005xa, Cliche:2009fma, Lin:2010zzb, MartinMartinez:2012sg, Salton:2014jaa, Martin-Martinez:2015qwa, Zhou:2017axh, Cai:2018xuo, Ng:2018ilp, Pan:2020tzf, Barman:2021oum, Tjoa:2022oxv}, which encompasses the field of relativity and quantum information, is getting increasingly more attention due to its ability to probe the quantum nature of physical systems. An important area of research within RQI involves the study of entanglement harvesting from the background quantum field coupled to suitable detectors \cite{VALENTINI1991321, Reznik:2002fz, Reznik:2003mnx, Salton:2014jaa}, one of the most prominent of them being the Unruh-deWitt detectors \cite{Crispino:2007eb}. These detectors were originally introduced to understand the Unruh radiation as perceived by an accelerated observer from the Minkowski vacuum \cite{Unruh:1976db}, and, subsequently, were used for studying Hawking radiation in black hole spacetimes \cite{hawking1975}. In entanglement harvesting, one usually considers two such Unruh-DeWitt detectors that are initially uncorrelated and then investigates the condition for these two detectors to get entangled over time. This detector entanglement depends on various scenarios, like due to the motion of the detectors \cite{Reznik:2002fz, Floreanini:2004, Salton:2014jaa, Koga:2018the, Koga:2019fqh, Zhang:2020xvo, Barman:2022xht, Suryaatmadja:2022quq}, background spacetime geometry \cite{FuentesSchuller:2004xp, Cliche:2010fi, MartinMartinez:2012sg, Henderson:2017yuv, Kukita:2017etu, Henderson:2018lcy, Robbins:2020jca, Tjoa:2020eqh, Cong:2020nec, Gallock-Yoshimura:2021yok, Barman:2021kwg, Tjoa:2022oxv, Barman:2023rhd, Xu:2020pbj, Gray:2021dfk}, presence of a thermal bath \cite{Brown:2013kia, Simidzija:2018ddw, Barman:2021bbw} to name a few among a plethora of other possibilities \cite{Pozas-Kerstjens:2015gta, Pozas-Kerstjens:2016rsh, Sachs:2017exo, Cong:2018vqx, Henderson:2020ucx, Stritzelberger:2020hde}. {  The experimental prospects of entanglement harvesting have been proposed in atomic physics \cite{Eduardo:2013}, with superconducting qubits \cite{Eduardo_prl:2012}, between forward and past light cones \cite{Olson:2011}. Utilizing the light-matter interaction analogy of Unruh-deWitt detectors \cite{Eduardo_udw:2013}, entanglement harvesting from the electromagnetic vacuum is also studied, and its experimental prospect is provided in \cite{Eduardo:2016}.}

{  In the existing literature on entanglement harvesting, studying its features in gravitational wave background has received considerable attention lately \cite{Xu:2020pbj, Gray:2021dfk}. Investigating entanglement in oscillatory GWs have been done in \cite{Xu:2020pbj}. However, GWs carrying energy and momentum fluxes  can cause permanent distortion in spacetime, known as the memory effect. It is realized as a permanent shift in the separation between test particles \cite{Favata:2010, Tolish_wald:2014}. Apart from radiative effects, gravitational memory is linked with asymptotic symmetries (also known as BMS symmetries) and soft theorems \cite{Strominger:2017}. Using this triangular relationship, there have been investigations  on {\em quantum memory effect} in Rindler  and Schwarschild spacetimes \cite{Kolekar_rindler:2017,Kolekar_Schw:2023}. In \cite{Kolekar_rindler:2017}, the authors have shown how the Bogoliubov coefficients have non-trivial mode-mixing before and after the passage of a BMS shock wave, and also how the Negativity measure of entanglement gets modified. In the present article, we try to measure the harvested entanglement by computing the concurrence for a GW burst spacetime. The spacetime geometry consists of linearized perturbations of GW burst propagating over a Minkowski background. To discern the relationship between memory and the concurrence measure on a more firm footing, we study both with and without memory GW burst profiles.}

{  Burst profiles studied in linearized gravity contain memory only when there is an asymmetry in the GW pulse profile.  Mathematically speaking, the asymptotic value of the metric perturbation  at infinite past and future are different \cite{Zeldovich:1974, Tolish_wald:2014}. Gaussian and sech-squared pulse profiles are chosen for bursts without memory. On the other hand, the profiles which contain GW memory are chosen to be Heaviside step function and tanh. The choice of Heaviside theta function is taken as it is used frequently for theoretical modelling \cite{Aurrekoetxea:2020,Wang:2014}. In addition to this profile we have chosen to work with a continuous tanh pulse profile because it captures the basic physics of memory and provides a sense of the duration of the burst.} Memory effects using different pulse profiles in exact solutions of radiative geometries have been extensively studied (see \cite{Zhang_memory,Chakraborty:2019,Chakraborty:2020,Chakraborty:2021}). On the phenomenological front, memory signals for binary mergers \cite{Favata:2010}, core-collapse supernovae \cite{Abbott:supernova}, hyperbolic scattering of flybys, gravitational bremsstrahlung \cite{Abbott:supernova, Kovacs:1978, Garcia-Bellido:2017}, and gamma-ray bursts \cite{Sago:2004} have also been worked out. Till date, memory effect has not been observed  as it is a relatively weak effect and beyond the scope of present LIGO detectors \cite{Hubner:2021}. However, with upcoming third generation detectors, there exists possibility of its detection \cite{Favata:2010,Grant:2022,Ghosh:2023}.

{  Coming back to the present work, our setup consists of two {\em static} Unruh-DeWitt detectors coupled to a quantum scalar field in the previously mentioned GW burst spacetimes. The main objective of the paper is to understand the features of entanglement harvesting due to gravitational memory. We address this  by comparing burst profiles with and without memory. Although both types of profiles show entanglement harvesting, there exist qualitative differences. We indeed find that bursts with memory show an inverse scaling between the detector energy gap and the concurrence measure in the low-energy limit. This is argued to be similar to the Weinberg leading term in the soft-graviton theorem \cite{Weinberg:1965,Strominger:2017}.} Furthermore, we investigate the characteristics of this harvested entanglement depending on different system parameters. We observe some generic features in the harvested entanglement in all these cases. For instance, harvesting decreases with increasing detector transition energy and the distance between the two detectors.  %We find that the harvesting profile, in low detector transition energy regimes, exhibits some distinct features for the Gaussian and sech-squared profiles, which are without asymptotic memory, compared to the tanh profile with asymptotic memory.
{  Since we work with static detectors, the motion of the detectors has little effect on harvesting, and the role of the spacetime geometry in determining the measure of the entanglement is noted. The difference in concurrence arises because the spacetime geometry
is different in the case of bursts with and without memory ({\em i.e.} the metric functions are different).} We elaborately discuss these differences, point out crucial entanglement harvesting-related features specific to the different scenarios, and discuss the reasoning and interpretation of our obtained results. We also compare our entanglement harvesting-related findings with previous results \cite{Xu:2020pbj} of periodic GW passing through flat spacetime. 

This manuscript is organized in the following manner. In Sec. \ref{sec:model-set-up}, we provide the basic formalism to understand the condition of entanglement harvesting with two Unruh-DeWitt detectors and elucidate the measure of the harvested entanglement concurrence. In Sec. \ref{sec:metric-Greens-fn} we consider the GW background of our choice and discuss the gravitational memory effect. Following in Sec. \ref{sec:scalar-field-WightmanFn} we construct the necessary Green functions with a massless minimally coupled scalar field to understand the entanglement harvesting in the considered background. In the subsequent Sec. \ref{sec:entanglement-harvesting}, we investigate the entanglement harvesting condition and study the concurrence. In this section, we also detail our observations on entanglement harvesting with the considered GW profiles.   {In Sec. \ref{sec:Detectors-trajectories-Minkowski}, we consider non-geodesic trajectories for the detectors in Minkowski spacetime that are inspired by the geodesic trajectory in the GW burst backgrounds and try to understand whether these detectors can produce similar harvesting profiles like the static detectors in GW backgrounds.} Finally, in Sec. \ref{sec:discussion}, we discuss the main observations and the consequence of our findings and provide our concluding remarks.

\section{Basic formalism and model}\label{sec:model-set-up}

In this section, we elucidate the model setup, which is comprised of  two Unruh-DeWitt detectors. These are point-like  two-level atomic detectors and were first conceptualized to understand the Unruh effect \cite{Unruh:1976db}. We consider a scenario where one of the detectors is carried by observer Alice. This detector is denoted by $A$. At the same time, the other detector carried by Bob is denoted by $B$. Furthermore, we consider the detector states $|E_{n}^{j}\rangle$ to be non-degenerate, with $j=A, B$ and $n=0,1$ denoting the individual detector and the excitations, respectively. These detectors are assumed to interact with a background massless scalar field $\Phi(X)$ through monopole interactions $m_{j}(\tau_{j})$. The entire interaction action for this two-detector system is given by 
\begin{eqnarray}\label{eq:TimeEvolution-int}
 \mathcal{S}_{int} &=& \int_{-\infty}^{\infty} c\,\bigg[ 
\kappa(\tau_{A})\,m_{A}(\tau_{A})\, \Phi\left(X_{ A}(\tau_{A})\right) 
d\tau_{A} \nonumber\\ 
&& +~ \kappa(\tau_{B})\,m_{B}(\tau_{B})\, \Phi\left(X_{ 
B}(\tau_{B})\right) d\tau_{B}\bigg]~,
\end{eqnarray}
where $c$ denotes the coupling strength of interaction between the detectors and the scalar field, which we have considered to be the same in the case of both detectors. However, this coupling can also be considered different with different detectors in a more general scenario. In the previous expression, the other quantities $\kappa_{j}(\tau_{j})$ and $\tau_{j}$ respectively denote the specific switching functions and the proper time corresponding to a particular detector. Moreover, we consider the detectors and the field to be initially in the ground state, i.e., in the state, $|in\rangle = |0\rangle |E_{0}^{A}\rangle |E_{0}^{B}\rangle$, where $|0\rangle$ denotes the field's ground state. Then the final state of the system is obtained from the time evolution of this state as $|out\rangle = T\left\{e^{i \mathcal{S}_{int}}|in\rangle\right\} $, with $T$ signifying the necessary time ordering. Then by tracing out the field states and treating the coupling strength $c$ between the detectors and the field in a perturbative manner, one can express the final reduced detector density matrix of the system as
\begin{equation}\label{eq:detector-density-matrix}
 \rho_{AB} = c^2\,
 {\left[\begin{matrix}
 0 & 0 & 0 & \varepsilon\\~\\
0 & P_{A} & P_{AB} & W_{A}\\~\\
0 & P_{AB}^{*} & P_{B} & W_{B}\\~\\
\varepsilon^{*} & W_{A}^{*} & W_{B}^{*} & 
1/c^2-(P_{A}+P_{B})
 \end{matrix}\right]}
 +\mathcal{O}(c^4)~.
\end{equation}
This density matrix is expressed on the basis of the field states $\big\{|E_{1}^{A}\rangle |E_{1}^{B}\rangle, |E_{1}^{A}\rangle |E_{0}^{B}\rangle, |E_{0}^{A}\rangle |E_{1}^{B}\rangle, |E_{0}^{A}\rangle |E_{0}^{B}\rangle\big\}$. The expressions of the quantities in this density matrix that will be relevant for understanding entanglement harvesting are 
\begin{subequations}\label{eq:Pj-epsilon}
\begin{eqnarray}
 P_{j} &=& |\langle E_{1}^{j}|m_{j}(0)|E_{0}^{j}\rangle|^2 \,
\mathcal{I}_{j}~,\\
\varepsilon &=& \langle E_{1}^{B}|m_{B}(0)|E_{0}^{B}\rangle\langle 
E_{1}^{A}|m_{A}(0)|E_{0}^{A}\rangle \,
\mathcal{I}_{\varepsilon}~.
\end{eqnarray}
\end{subequations}
We should mention that the other quantities $P_{AB}$ and $W_{j}$ do not contribute to the condition of entanglement harvesting or in the measure of the harvested entanglement \cite{Koga:2018the}. Therefore, we do not provide their explicit expressions here. These details can be found in Ref. \cite{Koga:2018the}. Finally, in the previous equation the quantities $\mathcal{I}_{j}$ and $\mathcal{I}_{\varepsilon}$ are given by 
\begin{subequations}\label{eq:all-integrals}
\begin{eqnarray}
 \mathcal{I}_{j} &=& \int_{-\infty}^{\infty}d\tau'_{j} 
\int_{-\infty}^{\infty}d\tau_{j}~e^{-i\Delta E^{j}(\tau'_{j}-\tau_{j})}\nonumber\\
~&& ~~~~~~\times~
\kappa(\tau'_{j})\kappa(\tau_{j})\,G_{W}(X'_{j},X_{j})~,\\
\mathcal{I}_{\varepsilon} &=& -i\int_{-\infty}^{\infty}d\tau'_{B} 
\int_{-\infty}^{\infty}d\tau_{A}~e^{i(\Delta 
E^{B}\tau'_{B}+\Delta E^{A}\tau_{A})} \nonumber\\
~&& ~~~~~~\times~\kappa(\tau'_{B})\kappa(\tau_{A})\,G_{F}(X'_{B},X_{A})~,
\end{eqnarray}
\end{subequations}
where the detector states are assumed to be non-degenerate $E_{1}^{j}\neq E_{0}^{j}$, and we have further assumed $\Delta E^{j} = E_{1}^{j}- E_{0}^{j}>0$. Here $G_{W}(X_{j},X_{j'})$ and $G_{F}(X_{j},X_{j'})$ respectively denote the positive frequency Wightman function with $X_{j}>X_{j'}$, and the Feynman propagator. Their expressions are
\begin{subequations}\label{eq:Greens-fn-gen}
\begin{eqnarray}\label{eq:Greens-fn-gen-W}
 G_{W}\left(X_{j},X_{j'}\right) &\equiv& \langle 
0_{M}|\Phi\left(X_{j}\right)\Phi\left(X_{j'}\right)|0_{M}
\rangle~,\\
 G_{F}\left(X_{j},X_{j'}\right) &\equiv& -i\langle 
0_{M}|T\left\{\Phi\left(X_{j}\right)\Phi\left(X_{j'}\right)\right\}|0_{M}
\rangle~\nonumber\\
~&=& -i\big[\theta(t_{j}-t_{j'})\,G_{W}(X_{j},X_{j'})\nonumber\\
~&& ~+~ \theta(t_{j'}-t_{j}) 
\,G_{W}\left(X_{j'},X_{j}\right)\big]~,\label{eq:Greens-fn-gen-F}
\end{eqnarray}
\end{subequations}
where $\theta(t)$ is the Heaviside theta function. Necessarily in these expressions, one can generally have $j\neq j'$, i.e., $j$ and $j'$ may correspond to two different detectors. Moreover, one should note that the expression of $\mathcal{I}_{j}$ contains only the index $j$, and the $X'_{j}$, $\tau'_{j}$ correspond to a different event and its proper time respectively, for the same detector. The world line $X_{j}$ of $j^{th}$ detector has components $X^{\mu}_{j}= (t_{j},x_{j},y_{j},z_{j})$.

According to \cite{Peres:1996dw, Horodecki:1996nc}, a general bipartite system can become entangled if the partial transposition of the corresponding density matrix has negative eigenvalues. With the reduced density matrix from Eq. (\ref{eq:detector-density-matrix}), this condition, see \cite{Koga:2018the}, results in
\begin{equation}\label{eq:cond-entanglement-harvesting}
 P_{A}P_{B}<|\varepsilon|^2~, 
\end{equation}
which, after removing the common monopole moment expectation values from both sides, looks like \cite{Koga:2018the, Koga:2019fqh}
\begin{equation}\label{eq:cond-entanglement}
 \mathcal{I}_{A}\mathcal{I}_{B}<|\mathcal{I}_{\varepsilon}|^2~.
\end{equation}
Note that using the expression of the Feynman propagator from Eq. (\ref{eq:Greens-fn-gen-F}), one can represent the integral $\mathcal{I}_{\varepsilon}$ as
\begin{eqnarray}\label{eq:Ie-integral}
 && \mathcal{I}_{\varepsilon} = -\int_{-\infty}^{\infty}d\tau_{B}  
\int_{-\infty}^{\infty}d\tau_{A}~\scalebox{0.91}{$e^{i(\Delta 
E^{B}\tau_{B}+\Delta E^{A}\tau_{A})} $} \kappa(\tau_B)\kappa(\tau_A)\times\nonumber\\
~&& \scalebox{0.95}{$\left[\theta(t_{B}-t_{A})\,G_{W}(X_{B},X_{A})+ \theta(t_{A}-t_{B}) 
\,G_{W}\left(X_{A},X_{B}\right)
\right].$}\nonumber\\
\end{eqnarray}
Thus, all quantities in (\ref{eq:cond-entanglement}), necessary for understanding the phenomenon of entanglement harvesting, can be obtained in terms of the Wightman functions.

Once the possibility of entanglement harvesting is established through the satisfaction of the condition (\ref{eq:cond-entanglement}), one can investigate different entanglement measures to quantify the harvested entanglement. Some of these measures are, see \cite{Zyczkowski:1998yd, Vidal:2002zz, Eisert:1998pz, Devetak_2005}, entanglement negativity, concurrence, etc. In particular, entanglement negativity is given by the sum of all negative eigenvalues of the partial transpose of $\rho_{AB}$, and signifies the upper bound of distillable entanglement. While in the case of a two-qubit system, the more convenient entanglement measure is concurrence $\mathcal{C}(\rho_{AB})$, see \cite{Koga:2018the, Koga:2019fqh, Hu:2015lda, Bennett:1996gf, Hill:1997pfa, Wootters:1997id}. In a two-qubit system \cite{Koga:2018the, Koga:2019fqh}, the concurrence is given by
\begin{eqnarray}\label{eq:concurrence-general}
    \mathcal{C}(\rho_{AB}) &=& max\Big[0,\, 2c^2 \left(|\varepsilon|-\sqrt{P_{A}P_{B}}\right)+\mathcal{O}(c^4)\Big]\nonumber\\
    ~&=& max\Big[0,\, 2\,c^2\,|\langle E_{1}^{B}|m_{B}(0)| E_{0}^{B}\rangle| 
 \nonumber\\
~&\times&
|\langle 
E_{1}^{A}|m_{A}(0)| E_{0}^{A}\rangle|  
\left(|\mathcal{I}_{\varepsilon}|-\sqrt{\mathcal{I}_{A}\mathcal{I}_{B}}
\right)+\mathcal{O} (c^4)\Big]~.\nonumber\\
\end{eqnarray}
The quantities $|\langle E_{1}^{j}|m_{j}(0)| E_{0}^{j}\rangle|$ are only dependent on the detectors' specific internal structure. They do not depend on the considered spacetime, motion of the detectors, and background scalar fields. Then as long as one wants to understand the effects of spacetime curvature or the motion of the detectors in the harvested entanglement, it is sufficient to study the quantity
\begin{equation}\label{eq:concurrence-I}
\mathcal{C}_{\mathcal{I}} = \left(|\mathcal{I} 
_{\varepsilon}| -\sqrt{\mathcal{I}_{A}\mathcal{I}_{B}} \right),
\end{equation} 
as a relevant contribution from the concurrence \cite{Koga:2018the, Koga:2019fqh}. We shall investigate $\mathcal{C}_{\mathcal{I}}$ in our considered system to talk qualitatively about the harvested entanglement.

%%%%%%%%%%%%%%%%%%%%%%%%%%%%%%%%%%%%%%%%%%%%%%%%%%%%%%%%%%%%%%%%%%%%%%%%%%%%%%%%%%%%%%%%%%%%%%%%%%%%%%%%%%%%%%%%%%%%%%%%%%%%%%%%%%%%%%%%%%%%%%%%%%%%%%
\section{Gravitational wave burst and the memory effect}\label{sec:metric-Greens-fn}

 The gravitational memory effect is a permanent change in a detector caused by the passing of a GW signal. Imagine two freely falling detectors (following geodesic trajectories) in the background gravitational field, and their separation is $d_{\rm initial}$. Now, assume that a GW signal interacts with the given system for a finite amount of time. After a significant amount of time has passed since the passage of GW signal, we further note the separation between these two detectors, say it to be $d_{\rm final}$. Turns out that $d_{\rm final} \neq d_{\rm initial}$, for a GW signal with memory and $d_{\rm final} = d_{\rm initial}$ whenever there exists no memory in the signal. This change in the relative separation in case of memory is expected to have measurable imprints \cite{Favata:2010}. In linearised gravity, this effect is simply realized by integrating the geodesic deviation equation. One finds after integration that the change in geodesic separation ($\xi^i$) is related to the change in metric perturbation ($h_{ij}$) of the GW, $\Delta\xi^i=\frac{1}{2}\Delta h^i\,_j\,\xi^j$ \cite{Tolish_wald:2014}. This difference in the metric perturbation at infinite past and future is the origin of memory \cite{Zeldovich:1974}. For non-relativistic sources, one identifies the memory term with the non-zero difference in the quadrupole moment of the source \cite{Favata:2010}.  
%
%
%
%
%\sajal{We do not need this} \sout{In this work, we are motivated to analyze the entanglement harvesting procedure for certain GW burst profiles with and without memory. We consider three burst profiles- i) Gaussian, ii) sech-squared iii) tanh. In the following Sec. \ref{sec:entanglement-harvesting}, we will show how the presence of memory in the tanh-burst profile has a significant impact on the nature of the harvested entanglement. }
%
%}\label{sec:Memory term}

%As mentioned earlier, GW bursts are generally predicted to be generated from events like hyperbolic scattering \cite{Garcia-Bellido:2017}, core-collapse supernovae \cite{Abbott:supernova}, gravitational bremsstrahlung \cite{Kovacs:1978}. Unlike the GW signal from periodic sources, the burst profiles are comparatively difficult to model, thus making the possibility of detection challenging \cite{Abbott:burst}. On the theoretical side, burst signals has been extensively studied in the context of memory effects (see \cite{Zhang:2016pqx}). Our work primarily broadens the scope of GW memory by establishing a correpsondence with entanglement measures between two Unruh-deWitt detectors.
 
To this end, we consider a toy-model burst scenario. The spacetime is comprised of linearized GW perturbations over Minkowski spacetime in the Transverse-Traceless (TT) gauge. The line element is,
\begin{eqnarray}\label{eq:BJR-metric}
ds^2 &=& -du\,dv+dx^2\,[1+f(u)]+dy^2\,[1-f(u)]\,.
\end{eqnarray}
%For plane GWs the two most well known coordinate systems are the Baldwin-Jeffery-Rosen (BJR) coordinates and the Brinkmann coordinates, see \cite{} and the references therein. In this work, we consider the BJR coordinates with GW burst like perturbation to the flat metric. The main reason behind this choice is that the scalar field equation of motion is easier to solve with the BJR metric compared to the Brinkmann. Thus from the point of view of investigating quantum field theoretic aspects, it is more suitable. Although there are some works dealing with these aspects in the Brinkmann coordinates \cite{}. We get additional motivation for considering the BJR coordinates from the fact that in \cite{} the authors have considered these coordinates to understand entanglement harvesting from a GW background that has periodic perturbations to the flat space. Then it becomes compelling to obtain our results in the BJR coordinates for comparing them with \cite{}. In particular, we consider the GW memory metric in BJR coordinates as
%
%
%
To arrive at the above expression, we assume that the GW is propagating in z-direction. The quantities $u$ and $v$ are defined as, $u=t-z$ and $v=t+z$, denoting the outgoing and ingoing null directions respectively. The metric in this form is pure 'plus' ($+$) polarization as there exist no cross terms (such as $dxdy$). Note that the contribution of memory effect coming from cross-polarization can be conveniently avoided by considering a suitable choice of tetrad \cite{Favata:2010,Johnson:2018}. 
%
\iffalse 
  {Moreover, even if one considers the cross-polarization, their contribution to the Wightman function vanishes for static detectors. Therefore, the presence of cross-polarization becomes redundant as we also choose static detectors for our concern.}
%
Note these Wightman functions are utilized to obtain single detector transition probabilities and entanglement measures \cite{}. \fi
%
As discussed earlier, we consider three burst profiles for $f(u)$ --- (i) Gaussian ($f(u)= \mathcal{A}\,e^{-u^2/\rho^2}$), (ii) sech-squared ($f(u)=\mathcal{A}\, sech^2(u/\varrho)$), (iii) Heaviside step function ($f(u)=\mathcal{A}\,\theta(u)$) and (iv) tanh ($f(u)=\mathcal{A}\,\{1+tanh(u/\lambda)\}$). These profiles denote initially and finally flat spacetimes, much akin to a burst. However, only in the last two cases, we find memory since the metric perturbation at initial and final times are different. Qualitatively similar burst profiles were obtained in the case of gravitational bremsstrahlung in \cite{Kovacs:1978}. Note that in the linearized regime $\mathcal{A}<<1$, the metric is a vacuum solution (only one of the Ricci tensors scales as $\mathcal{A}^2$, and others vanish). When $\mathcal{A}$ does not satisfy this constraint, the metric is a nonlinear, exact solution of general relativity and is known as exact plane waves  \cite{Griffiths:2009}.\footnote{The metric Eq.(\ref{eq:BJR-metric}) is the linearized form of the more general exact plane wave expressed in Baldwin-Jeffrey-Rosen (BJR) coordinates.} For exact plane waves, the same metric can be sourced by massless radiation like electromagnetic waves or neutrinos \cite{Griffiths:2009}.

Only one of the GW burst profile $(\tanh$-pulse) studied in this paper contains memory-type signal. We clarify this issue in this section. Let us rewrite Eq. (\ref{eq:BJR-metric}) in the following way, 
\begin{equation} \label{eq:perturbative}
    ds^2=\eta_{\mu\nu} dx^\mu\,dx^\nu+h_{\mu\nu } dx^\mu\,dx^\nu, 
\end{equation}
where Greek indices run over all the spacetime coordinates $(u,v,x,y)$, and the perturbation, $h_{\mu \nu}$, has the following form:
%%%%
\begin{widetext}
%%%%%%%%%%%%%%%%%%%%%%%
\begin{equation}
h_{\mu \nu} = 
\begin{bmatrix}
0 & 0 & 0 & 0 \\
0 & 0 & 0 & 0 \\
0 & 0 & f(u) & 0  \\
0 & 0 & 0 & -f(u)
\label{eq:h_mu_nu}
\end{bmatrix}
\end{equation}
%%%%%%%%%%%%%%%%%%%%%%%%
\end{widetext}
%%%%%%%%%%%%%%%%%%%%%%%%
%%%%
%The perturbation is, $h_{\mu\nu}=Diag [0,0,f(u),-f(u)]$. 
%
%Consider two geodesic timelike observers, whose 4-vector is given by $U^\mu$, traversing spacetime given in Eq. (\ref{eq:perturbative}). The geodesic deviation equation reads,
%
In order to invoke the concept of GW memory, we start with the \textit{geodesic deviation} equation \cite{book:wald}. Assume that we have a family of closely spaced geodesics, each parameterized with $\tau$, and the 4-vector for a particular geodesic is given as $X^{\mu}=dx^{\mu}/d\tau$. Let us define another variable $s$, which index different geodesics, and the connecting vector (also known as the deviation vector), $\xi^{\mu}$ reads as $\xi^{\mu}=dx^{\mu}/ds$. The governing equation to relate $\xi^{\mu}$ with the spacetime curvature is given as
\begin{equation} \label{eq:deviation_gen}
    \dfrac{D^2\xi^{\mu}}{d\tau^2}=\ddot{\xi}^{\mu}=- R^\mu\,_{\alpha\beta\gamma} X^\alpha \xi^\beta X^\gamma,
\end{equation}
For the spacetime metric given in Eq. (\ref{eq:perturbative}), the Riemann tensor turns out to be
\begin{equation}
    R^{\mu}\,_{\alpha\beta\gamma}=\frac{1}{2} (\partial_\alpha \partial_\beta h^\mu\,_\gamma + \partial_\gamma \partial^\mu h_{\alpha\beta} - \partial_\beta \partial_\gamma  h^\mu\,_{\alpha} - \partial_\alpha \partial^\mu  h_{\beta\gamma} ).
\end{equation}
Note that we are working within the weak field approximation, as evident from the fact that Riemann components only contain terms proportional to $f(u)$. Within this domain, we can safely assume the 4-velocity to be timelike, and $X^{\mu}=(1,0,0,0)$. Finally, geodesic deviation equation (\ref{eq:deviation_gen}) becomes
%
%Since the Riemann tensor already contain terms proportional to $f(u)$, the zeroth order term for the 4-velocity can be taken as, $U^\mu=(1,0,0,0)$. Considering TT gauge, in the linear order, the geodesic deviation equation (\ref{eq:deviation_gen}) becomes,
%%%
\begin{equation}\label{eq:deviation_memory}
\Ddot{\xi^i}=-R^i\,_{uju}\xi^j=\frac{1}{2}\ddot{ h}^i\,_j \xi^j,
\end{equation}
%%%
where the Latin indices denote the spatial components ($x,y$).  The above expression can be solved in an iterative fashion by assuming that the change in separation caused by the GW burst is small compared to the original separation. Let the separation vector be expressed as, $\xi^i(u)=\xi^i_0+\Delta \xi^i(u)$, where $\xi^i_0$ is the original separation, and $\Delta \xi^i(u)$ is the change in separation. Now, in the linear order, we get
%%%
\begin{equation}\label{eq:memory_main}
    \Delta\xi^i=\frac{1}{2}\Delta h^i\,_j \xi_0^j\simeq \frac{1}{2}\Delta h^i\,_j \xi^j.
\end{equation}
%%%
The above equation is the main crux of the GW memory effect \cite{Tolish_wald:2014, book:Schutz, maggiore2007gravitational}.  Therefore, in linearized theory, to observe memory in a Minkowski background, the metric perturbation of the GW at asymptotic future and past should differ. In the burst profiles, the asymptotic value at $u\to\pm\infty$ is the same for both the Gaussian and sech-squared pulse, as they are symmetric. Since the step function and tanh pulses are asymmetric, the value at the two asymptotic limits differ; hence, we find a memory effect. Later we will see how this feature of the burst profiles has consequences for entanglement harvesting.

%%%%%%%%%%%%%%%%%%%%%%%%%%%%%%%%%%%%%%%%%%%%%%%%%%%%%%%%%%%%%%%%%%%%%%%%%%%%%%%%%%%%%%%%%%%%%%%%%%%%%%%%%%%%%%%%%%%%%%%%%%%%%%%%%%%%%%%%%%%%%%%%%%%%%%

\section{Propagation of the scalar field and the Wightman function}\label{sec:scalar-field-WightmanFn}

Now, let us consider a massless minimally coupled free scalar field $\Phi(X)$ in the previously mentioned spacetime. The corresponding equation of motion $\Box \Phi(X) = (1/\sqrt{-g})\,\partial_{\mu} \big[\sqrt{-g}\,g^{\mu\nu}\,\partial_{\nu}\Phi\big]=0$, in the background of Eq. (\ref{eq:BJR-metric}) takes the form 
\begin{eqnarray}\label{eq:scalar-field-EOM-1}
    -2\,\partial_{u}\partial_{v}\Phi +\frac{1}{2}\bigg[\frac{\partial_{x}^2}{1+f(u)} + \frac{\partial_{y}^2}{1-f(u)}\bigg]\,\Phi = 0~.
\end{eqnarray}
To solve this differential equation, one may consider the field decomposition $\Phi \sim \mathcal{R}(u)\times\exp{\{i(-k_{-}v+k_{1}x+k_{2}y)\}}$ \cite{Xu:2020pbj, Garriga-1991}, as the spacetime is symmetric under the translation in $x$, $y$, and $v$. Then by finding the expression of $\mathcal{R}(u)$ utilizing Eq. (\ref{eq:scalar-field-EOM-1}), and with a suitable definition of the inner product among the modes \cite{Xu:2020pbj}, one can construct normalized mode functions \cite{Xu:2020pbj, Garriga-1991}. Let us consider these orthonormal modes in general to be $u_{\mathbf{k}}(X)$. Introducing annihilation and creation operators corresponding to the positive and negative frequency field modes, one can decompose the scalar field \cite{book:Birrell} as
\begin{eqnarray}\label{eq:scalar-field-decomp}
\Phi(X) = \int\,d^3k\,\big[u_{\mathbf{k}}(X)\,\hat{a}_{\mathbf{k}}+u^{\star}_{\mathbf{k}}(X)\,\hat{a}^{\dagger}_{\mathbf{k}}\big]~.
\end{eqnarray}
Here the annihilation and the creation operators satisfy the commutation relation $[\hat{a}_{\mathbf{k}},\hat{a}^{\dagger}_{\mathbf{k}'}] = (2\pi)^3\delta(\mathbf{k}-\mathbf{k}')$, and the operator $\hat{a}_{\mathbf{k}}$ annihilates the vacuum, say $|0\rangle$. Using this commutation relation and the mode expansion from Eq. (\ref{eq:scalar-field-decomp}), one can find out the Wightman function as $\langle 0|\,\Phi(X)\,\Phi(X')\,|0\rangle = \int\,d\mathbf{k} \,u_{\mathbf{k}}(X)\,u^{\star}_{\mathbf{k}}(X')$.   {For very small gravitational wave perturbation strength, i.e., $\mathcal{A}\ll1$, one may obtain the Wightman function by perturbative expansion as}
\begin{eqnarray}\label{eq:Wightman-fn-general}
    G_{W}(X,X') = G_{W_{M}}(X,X') + G_{W_{GW}}(X,X')~.
\end{eqnarray}
  {Here, $G_{W_{M}}(X,X')$ and $G_{W_{GW}}(X,X')$ respectively correspond to the contributions solely due to the flat Minkowski spacetime and the GW (taking terms up to $\mathcal{O}(\mathcal{A})$) in the Wightman function.} In our following study, we shall first provide the actual expressions for these $G_{W_{M}}(X,X')$ and $G_{W_{GW}}(X,X')$ for the choice of a certain $f(u)$ as discussed previously.

%\subsection{Evaluation of the Wightman function}

Let us now evaluate the Wightman function corresponding to the field vacuum with the help of the field decomposition given in Eq. (\ref{eq:scalar-field-decomp}). It is to be noted that the Wightman function can be cast into a form (\ref{eq:Wightman-fn-general}) by perturbatively treating $\mathcal{A}$, like done in \cite{Xu:2020pbj}, corresponding to each of the considered GW burst profiles. These Wightman functions will then be utilized in the next section to estimate the concurrence according to Eq. (\ref{eq:concurrence-I}). Let us evaluate the Wightman functions corresponding to the different considered burst profiles one by one.

\subsection{When $f(u)=\mathcal{A}\, e^{-u^2/\rho^2}$}

Let us first consider a spacetime described by the metric in Eq. (\ref{eq:BJR-metric}) where $f(u)=\mathcal{A}\, e^{-u^2/\rho^2}$, i.e., with a Gaussian wave burst. In this background, we are going to estimate the expression of the Wightman function given by Eq. (\ref{eq:Wightman-fn-general}). In this regard, let us first start getting the normalized wave modes $u_{\mathbf{k}}(X)$. The general expression of this normalized field mode in an exact plane wave metric background is provided in \cite{Xu:2020pbj, Garriga-1991}. In the case of $f(u)=\mathcal{A}\, e^{-u^2/\rho^2}$, let us express these normalized field modes in an approximate manner as
\begin{eqnarray}\label{eq:mode-fn-Gaussian}
u_{\mathbf{k}}(X) &\simeq& \frac{1}{\sqrt{2k_{-}(2\pi)^3}}\,e^{-i\,k_{-}v+i\,k_{1}x+i\,k_{2}y-i\frac{(k_{1}^2+k_{2}^2)}{4k_{-}}u} \nonumber\\
~&\times& \exp{\bigg[\frac{i\,\mathcal{A}}{8k_{-}}(k_{1}^2-k_{2}^2)\Big\{\rho\sqrt{\pi} \, \erf\Big(\frac{u}{\rho}\Big) \Big\}\bigg]}\,.\nonumber\\
\end{eqnarray}
Since we work with a linearized gravitational wave solution, so in all future occurences we assume the condition $\mathcal{A}<<1$. Using this expression for the field mode and with the field decomposition from Eq. (\ref{eq:scalar-field-decomp}) one can obtain the specific parts of the Wightman function as specified in Eq. (\ref{eq:Wightman-fn-general}) as
\begin{subequations}\label{eq:Wightman-fn-Gaussian-1}
\begin{eqnarray}
    G_{W_{M}}(X,X') &=& -\frac{i}{4\pi^2\Delta u}\int_{0}^{\infty}dk_{-}\,e^{i\,k_{-}(\sigma_{M}/\Delta u)}~,\\
    G_{W_{GW}}(X,X') &=& \frac{\mathcal{A}(\Delta x^2-\Delta y^2)}{8\pi^2\Delta u^3}\,\bigg[\rho\sqrt{\pi}\Big\{\erf\Big(\frac{u}{\rho}\Big)\nonumber\\
    ~&-& \erf\Big(\frac{u'}{\rho}\Big)\Big\}\bigg]\, \int_{0}^{\infty}dk_{-}\,k_{-}\,e^{i\,k_{-}(\sigma_{M}/\Delta u)}~.\nonumber\\
\end{eqnarray}
\end{subequations}
Here $\sigma_{M}\equiv -\Delta u\,\Delta v+\Delta x^2+\Delta y^2$, i.e., the square of the Minkowski geodesic distance, and $\erf(x)$ signifies the error function. The integrals involved in the expression (\ref{eq:Wightman-fn-Gaussian-1}) are oscillatory and are formally divergent in the specified integration limits. However, introducing a multiplicative regulator of the form $e^{-k_{-}\epsilon}$ we can evaluate the above integrals as $\int_{0}^{\infty}dk_{-}\,e^{i\,k_{-}(\sigma_{M}/\Delta u)}\, e^{-k_{-}\epsilon} =  (i)/(\sigma_{M}/\Delta u+i\,\epsilon)$ and $\int_{0}^{\infty}dk_{-}\,k_{-}\,e^{i\,k_{-}(\sigma_{M}/\Delta u)}\, e^{-k_{-}\epsilon} =  -1/(\sigma_{M}/\Delta u+i\,\epsilon)^2$. Thus the expressions of the previous Wightman function components can be obtained as
\begin{subequations}\label{eq:Wightman-fn-Gaussian-2}
\begin{eqnarray}
    G_{W_{M}}(X,X') &=& \frac{1}{4\pi^2\Delta u}\times\frac{1}{\sigma_{M}/\Delta u+i\,\epsilon}~,\label{eq:Wightman-fn-Gaussian-2-M}\\
    G_{W_{GW}}(X,X') &=& -\frac{\mathcal{A}(\Delta x^2-\Delta y^2)}{8\pi^2\Delta u^3}\, \frac{1}{(\sigma_{M}/\Delta u+i\,\epsilon)^2}\nonumber\\
    ~&& ~\times\,\bigg[\rho\sqrt{\pi}\Big\{\erf\Big(\frac{u}{\rho}\Big) -\erf\Big(\frac{u'}{\rho}\Big)\Big\}\bigg]~.\nonumber\\\label{eq:Wightman-fn-Gaussian-2-GW}
\end{eqnarray}
\end{subequations}
We shall use these expressions for the Wightman function to obtain the integrals $\mathcal{I}_{j}$ and $\mathcal{I}_{\varepsilon}$, necessary for quantifying the measure of the harvested entanglement.

\subsection{When $f(u)=\mathcal{A}\, sech^2(u/\varrho)$}

Let us now consider the GW burst with $f(u)=sech^2(u/\varrho)$ in Eq. (\ref{eq:BJR-metric}). In this background, the equation of motion for a massless minimally coupled scalar field admits the mode solutions of the form given by:
\begin{eqnarray}\label{eq:mode-fn-sech2}
u_{\mathbf{k}}(X) &\simeq& \frac{1}{\sqrt{2k_{-}(2\pi)^3}}\,e^{-i\,k_{-}v+i\,k_{1}x+i\,k_{2}y-i\frac{(k_{1}^2+k_{2}^2)}{4k_{-}}u} \nonumber\\
~&\times& \exp{\bigg[\frac{i\,\mathcal{A}}{4k_{-}}(k_{1}^2-k_{2}^2)\,\varrho \, \tanh{\Big(\frac{u}{\varrho}\Big)} \bigg]}\,.\nonumber\\
\end{eqnarray}
Now one can decompose the scalar field as done in Eq. (\ref{eq:scalar-field-decomp}) with the help of these mode functions. Furthermore, one can express the Witghman function in a similar manner as a sum of the Minkowski and the GW contributions as done in Eq. (\ref{eq:Wightman-fn-general}). Thus now the $W_{GW}(x,x')$ term has new expression given by
\begin{eqnarray}\label{eq:Wightman-fn-sech2-1}
    G_{W_{GW}}(X,X') &=& -\frac{\mathcal{A}(\Delta x^2-\Delta y^2)}{4\pi^2\Delta u^3}\,\frac{1}{(\sigma_{M}/\Delta u+i\,\epsilon)^2}\nonumber\\
    ~&& ~\times\,\bigg[\varrho\,\Big\{\tanh{\Big(\frac{u}{\varrho}\Big)}-\tanh{\Big(\frac{u'}{\varrho}\Big)}\Big\}\bigg] ~,\nonumber\\
\end{eqnarray}
where $\sigma_{M}\equiv -\Delta u\,\Delta v+\Delta x^2+\Delta y^2$ is the square of the Minkowski geodesic distance. It is to be noted that the quantity $G_{W_{M}}(X,X')$, denoting the contribution in the Wightman function entirely from the flat space, is same for both the considered $f(u)$. Therefore, its form can be recalled from Eq. (\ref{eq:Wightman-fn-Gaussian-2}).

\subsection{When $f(u)=\mathcal{A}\,\theta(u)$}

We consider scalar field mode solutions in a background with burst profile $f(u)=\mathcal{A}\,\theta(u)$. The characteristics of burst profiles with memory are generalized by this $\theta(u)$ function type perturbations, see \cite{mohanty2023gravitational}. One can obtain the mode solution from Eq. (\ref{eq:scalar-field-EOM-1}) with this Heaviside theta function perturbation as:
\begin{eqnarray}\label{eq:mode-fn-thetaFn}
u_{\mathbf{k}}(X) &\simeq& \frac{1}{\sqrt{2k_{-}(2\pi)^3}}\,e^{-i\,k_{-}v+i\,k_{1}x+i\,k_{2}y-i\frac{(k_{1}^2+k_{2}^2)}{4k_{-}}u} \nonumber\\
~&\times& \exp{\bigg[\frac{i\,\mathcal{A}}{4k_{-}}(k_{1}^2-k_{2}^2)\,\,u\,\theta(u) \bigg]}\,.\nonumber\\
\end{eqnarray}
With the help of these mode functions, we decompose the scalar field as prescribed in Eq. (\ref{eq:scalar-field-decomp}). As discussed previously, when $\mathcal{A}\ll 1$, the Wightman function is expressed in the form (\ref{eq:Wightman-fn-general}) where the Minkowski part $G_{W_{M}}(X,X')$ is independent of the considered burst profile $f(u)$ and remains the same in all cases. Whereas the GW part in the Wightman function $G_{W_{GW}}(X,X')$ is now given by
\begin{eqnarray}\label{eq:Wightman-fn-thetaFn-1}
     G_{W_{GW}}(X,X') &=& -\frac{\mathcal{A}(\Delta x^2-\Delta y^2)}{4\pi^2\Delta u^3}\,\frac{1}{(\sigma_{M}/\Delta u+i\,\epsilon)^2}\nonumber\\
    ~&& ~\times\bigg[u\,\theta(u)-u'\,\theta(u')\bigg] ~.
\end{eqnarray}
\color{black}

\subsection{When $f(u)=\mathcal{A}\,\{1+tanh(u/\lambda)\}$}

Finally, let us now look into the scalar field mode solutions in a background with a GW burst profile $f(u)=\mathcal{A}\,\{1+tanh(u/\lambda)\}$. As discussed in sec. (\ref{sec:metric-Greens-fn}), this metric will result in a nonvanishing memory term. The scalar field satisfies the same equation of motion of Eq. (\ref{eq:scalar-field-EOM-1}). With $f(u)=\mathcal{A}\,\{1+tanh(u/\lambda)\}$, the mode solutions become:
\begin{eqnarray}\label{eq:mode-fn-tanh}
u_{\mathbf{k}}(X) &\simeq& \frac{1}{\sqrt{2k_{-}(2\pi)^3}}\,e^{-i\,k_{-}v+i\,k_{1}x+i\,k_{2}y-i\frac{(k_{1}^2+k_{2}^2)}{4k_{-}}u} \nonumber\\
~&\times& \exp{\bigg[\frac{i\,\mathcal{A}}{4k_{-}}(k_{1}^2-k_{2}^2)\,\Big\{u+\lambda \, \ln {\Big[\cosh{\Big(\frac{u}{\lambda}\Big)}\Big]}\Big\} \bigg]}\,.\nonumber\\
\end{eqnarray}
Again one can decompose the scalar field as done in Eq. (\ref{eq:scalar-field-decomp}) with the help of these mode functions and express the Wightman function as a sum of the Minkowski and the GW contributions like done in Eq. (\ref{eq:Wightman-fn-general}). The Minkowski part remains the same for any form of considered $f(u)$, and its form can be recalled from Eq. (\ref{eq:Wightman-fn-Gaussian-2}). The GW part $W_{GW}(X,X')$ is now given by
\begin{eqnarray}\label{eq:Wightman-fn-tanh-1}
    && G_{W_{GW}}(X,X') = -\frac{\mathcal{A}(\Delta x^2-\Delta y^2)}{4\pi^2\Delta u^3}\,\frac{1}{(\sigma_{M}/\Delta u+i\,\epsilon)^2}\nonumber\\
    ~&& ~\times\bigg[\varrho\,\Big\{\Delta u+\lambda \, \Big(\ln {\Big[\cosh{\Big(\frac{u}{\lambda}\Big)}\Big]}-\ln {\Big[\cosh{\Big(\frac{u'}{\lambda}\Big)}\Big]}\Big)\Big\}\bigg] ~.\nonumber\\
\end{eqnarray}
Here also, $\sigma_{M}$ denotes the square of the Minkowski geodesic distance, as mentioned earlier.

\section{Entanglement harvesting: Concurrence}\label{sec:entanglement-harvesting}

In this section, we consider the linearized gravitational wave spacetime as discussed in Sec. \ref{sec:metric-Greens-fn}, and investigate the entanglement harvesting condition and the measure of the harvested entanglement, i.e., we study the concurrence. To complete this task, we will need the expression of the necessary Wightman function, as evaluated before in the same section. Moreover, we consider both of our detectors to be static. The reason behind considering static detectors is that they are the simplest detector configuration, and in this case, one can analytically pursue some of the calculations. Moreover, it is observed \cite{Koga:2018the} that with static detectors in flat spacetime, one cannot harvest any entanglement. Thus, compared to the flat background, the static detectors can provide the simplest way to understand the effects of GWs in the entanglement harvesting profile. In particular, the trajectories of the detectors carried by \emph{Alice} and \emph{Bob} are considered to be $X^{\mu}_{A}=(t_{A},0,0,0)$ and $X^{\mu}_{B}=(t_{B},d,0,0)$. As is visible, the detectors are separated by a distance $d$ in the $x$ direction. Moreover, we have $\tau_{j}=t_{j}$ as the detectors are static. In our following analysis, we shall first evaluate the integrals $\mathcal{I}_{j}$ that correspond to individual detector transition probabilities and signify a local term in the concurrence given by Eq. (\ref{eq:concurrence-I}). After that, we shall evaluate $\mathcal{I}_{\varepsilon}$, which signifies the correlation between the two detectors and denotes a non-local contribution in the concurrence.

\subsection{Evaluation of $\mathcal{I}_{j}$}

Here we consider evaluating the local terms $\mathcal{I}_{j}$ in the entanglement measure concurrence (\ref{eq:concurrence-I}), which also signifies the transition probability of the $j^{th}$ detector. For this purpose, we consider the Wightman functions from Eq. (\ref{eq:Wightman-fn-general}) and (\ref{eq:Wightman-fn-Gaussian-1}), with Gaussian switching of the form $\kappa(\tau_{j})=e^{-\tau^2_{j}/(2\sigma^2)}$ and eternal switching $\kappa(\tau_{j})= 1$ respectively. Note that finite time switching, like the Gaussian one, is suitable from the point of view that one can specify the interval of interaction between the detectors and the field through these window functions. In the current scenario, the Gaussian switching, in particular, also allows us to compare the results with the previously obtained ones \cite{Xu:2020pbj} for periodic GWs. However, getting the expressions of $\mathcal{I}_{j}$ and $\mathcal{I}_{\varepsilon}$ using the Gaussian switching is not always possible for each of the considered burst profiles. Therefore sometimes it is convenient to consider the trivial eternal switching $\kappa(\tau_{j})= 1$. Moreover, as we shall see with this $\kappa(\tau_{j})= 1$ switching, all the flat space contributions in $\mathcal{I}_{j,\,\varepsilon}$ vanish. Therefore, it renders the entanglement measure free of any effects from non-trivial switching, and the concurrence depends only on the contribution from the GWs.
It should also be mentioned that the part $G_{W_{GW}}(X,X')$ of the Wightman function of Eq. (\ref{eq:Wightman-fn-general}), which originates completely due to the presence of the GW, has a quantity $(\Delta x^2-\Delta y^2)$ multiplied with it. In the present setup, the individual detectors are static, and therefore, this  quantity should vanish. Hence, in this case, the contribution of GWs in the single detector transition should vanish, which reproduces the findings of \cite{Gibbons:1975, Xu:2020pbj}. In particular, with a change of variables to $\eta=\tau'_{j}-\tau_{j}$ and $\xi=\tau'_{j}+\tau_{j}$ and with the Gaussian switching $\kappa(\tau_{j}) = e^{-\tau^2_{j}/(2\sigma^2)}$, the integral $\mathcal{I}_{j}$ from Eq. (\ref{eq:all-integrals}) would look like this:
\begin{eqnarray}\label{eq:Ij-Gaussian-1}
    \mathcal{I}_{j} &=& \frac{1}{2}\int_{-\infty}^{\infty}d\eta\int_{-\infty}^{\infty}d\xi\,e^{-\frac{(\eta^2+\xi^2)}{4\sigma^2}}e^{-i\,\Delta E\,\eta}\, G_{W_{M}}(X'_{j},X_{j})\nonumber\\
    ~&\simeq& -\frac{\sigma\sqrt{\pi}}{4\pi^2}\int_{-\infty}^{\infty}\frac{d\eta}{(\eta-i\epsilon/2)^2}\,e^{-\frac{\eta^2}{4\sigma^2}-i\,\Delta E\,\eta}~.
\end{eqnarray}
Here we have used the expression of $G_{W_{M}}(X,X')$ from Eq. (\ref{eq:Wightman-fn-Gaussian-2}), carried out the $\xi$ integration first using the Gaussian integration formula $\int_{-\infty}^{\infty}d\xi\,e^{-\alpha\,\xi^2}=\sqrt{\pi/\alpha}$, and considered $\epsilon$ to be a very small positive parameter to arrive at the final expression. Moreover, to carry out the integral in Eq. (\ref{eq:Ij-Gaussian-1}), one should express it using the Fourier transform of $e^{-\eta^2/(4\sigma^2)}$ as
\begin{eqnarray}\label{eq:Ij-Gaussian-2}
    \mathcal{I}_{j} &=& -\frac{\sigma^2}{4\pi^2}\int_{-\infty}^{\infty}d\zeta\,e^{-\zeta^2\sigma^2}\int_{-\infty}^{\infty}\frac{d\eta}{(\eta-i\epsilon/2)^2}\,e^{i(\zeta-\Delta E)\eta}~.\nonumber\\
    ~&=& \frac{\sigma^2}{2\pi}\int_{\Delta E}^{\infty}d\zeta\,(\zeta-\Delta E)\,e^{-(\zeta-\Delta E)\epsilon-\zeta^2\sigma^2}~.
\end{eqnarray}
The integral on $\eta$ in the second last step was nonzero only when $\zeta> \Delta E$. Therefore, the integration limit was changed appropriately. Now one can easily perform the integral in the last line, see \cite{Sriramkumar1996FinitetimeRO, Xu:2020pbj}, which in the limit of $\epsilon\to 0$ results in 
\begin{eqnarray}\label{eq:Ij-Gaussian-3}
    \mathcal{I}_{j} &=& \frac{1}{4\pi}\Big[e^{-\sigma^2\Delta E^2}-\sqrt{\pi}\sigma\Delta E\,\erf\text{c}(\sigma\,\Delta E)\Big]~,
\end{eqnarray}
where, $\erf\text{c}(x)\equiv 1-\erf(x) = (2/\sqrt{\pi})\int_{x}^{\infty}d\zeta\,e^{-\zeta^2}$ signifies the \emph{complementary error function} \cite{gradshteyn2007}. On the other hand, if we had used eternal switching, i.e., $\kappa(\tau_{j})=1$, then the expression of the individual detector transition probability $\mathcal{I}_{j}$ would have been
\begin{eqnarray}\label{eq:Ij-Infinite-1}
    \mathcal{I}_{j} &=& -\frac{1}{2}\int_{-\infty}^{\infty}d\xi\int_{-\infty}^{\infty}\frac{d\eta}{(\eta-i\epsilon/2)^2}\,e^{-i\,\Delta E\,\eta}~.
\end{eqnarray}
For $\Delta E>0$, we need to consider the contour in the lower half complex plane of $\eta$, to dampen the complexified integral. This does not contain any pole; thus, the integral has a vanishing contribution. Therefore, with an eternal switching function $\kappa(\tau_{j})=1$ we have the integral $\mathcal{I}_{j}=0$. This result is consistent with the usual notion that a static detector interacting for an infinite time with a background field in a flat spacetime will not detect any particle \cite{Koga:2018the}.

\subsection{Evaluation of $\mathcal{I}_{\varepsilon}$}

As we have seen, the Wightman function $G_{W}(X_{j},X_{j'})$ can be thought of to be a combination of two distinct terms. One is purely due to the Minkowski background ($G_{W_{M}}(X_{j},X_{j'})$), and another one is solely due to the GW ($G_{W_{GW}}(X_{j},X_{j'})$). Then utilizing this observation, the entire integral $\mathcal{I}_{\varepsilon}$ from (\ref{eq:Ie-integral}) denoting the entangling term can also be expressed as a sum $\mathcal{I}_{\varepsilon} = \mathcal{I}_{\varepsilon}^{M} + \mathcal{I}_{\varepsilon}^{GW}$, where these quantities are expressed as
\begin{widetext}
\begin{eqnarray}\label{eq:Ie-general-decomp}
    \mathcal{I}_{\varepsilon}^{M} &=& -\int_{-\infty}^{\infty}d\tau_{A}\int_{-\infty}^{\infty}d\tau_{B}\,\kappa(\tau_{A})\kappa(\tau_{B})\,e^{i\,\Delta E(\tau_{A}+\tau_{B})} \Big[\theta(t_{B}-t_{A})\,G_{W_{M}}(X_{B},X_{A})+\theta(t_{A}-t_{B})\,G_{W_{M}}(X_{A},X_{B})\Big]~,\nonumber\\
    \mathcal{I}_{\varepsilon}^{GW} &=& -\int_{-\infty}^{\infty}d\tau_{A}\int_{-\infty}^{\infty}d\tau_{B}\,\kappa(\tau_{A})\kappa(\tau_{B})\,e^{i\,\Delta E(\tau_{A}+\tau_{B})} \Big[\theta(t_{B}-t_{A})\,G_{W_{GW}}(X_{B},X_{A})+\theta(t_{A}-t_{B})\,G_{W_{GW}}(X_{A},X_{B})\Big]~.\nonumber\\
\end{eqnarray}
\end{widetext}
 We shall, one by one, evaluate these quantities in our following study.

\subsubsection{Evaluation of $\mathcal{I}_{\varepsilon}^{M}$}

To evaluate $\mathcal{I}_{\varepsilon}^{M}$, which corresponds to the contribution in the non-local entangling term only due to the Minkowski background, we first consider the expression from Eq. (\ref{eq:Ie-general-decomp}). It is to be noted that in this expression, the Wightman functions utilized correspond to two different detector points. Namely, it correlates detector $A$ with detector $B$ or vice versa. In this case, let us consider $\Bar{\eta}=\tau_{B}-\tau_{A}$, and $\Bar{\xi}=\tau_{B}+\tau_{A}$. Then one has $\Delta u_{BA}=u_{B}-u_{A}=\Bar{\eta}=-\Delta u_{AB}$, and also $\sigma_{M}=-\Delta t^2+\Delta \textbf{X}^2 = -\Bar{\eta}^2+d^2$. Finally, with the Gaussian window function $\kappa(\tau_{j})=e^{-\tau_{j}^2/(2\sigma^2)}$, one may express $\mathcal{I}_{\varepsilon}^{M}$, with Eq. (\ref{eq:Wightman-fn-Gaussian-2}), as
\begin{widetext}
\begin{eqnarray}\label{eq:Ie-M-Gaussian-1}
    \mathcal{I}_{\varepsilon}^{M} &=& \frac{1}{2}\int_{-\infty}^{\infty}d\Bar{\eta}\int_{-\infty}^{\infty}d\Bar{\xi}\,e^{-\frac{\Bar{\eta}^2+\Bar{\xi}^2}{4\sigma^2}+i\Delta E\,\Bar{\xi}}\times\frac{1}{4\pi^2}\bigg[\frac{\theta(\Bar{\eta})}{\Bar{\eta}^2-d^2-i\,\Bar{\eta}\,\epsilon}+\frac{\theta(-\Bar{\eta})}{\Bar{\eta}^2-d^2+i\,\Bar{\eta}\,\epsilon}\bigg],\nonumber\\
    ~&=& \frac{1}{4\pi^2}\int_{-\infty}^{\infty}d\Bar{\eta}\int_{-\infty}^{\infty}d\Bar{\xi}\,e^{-\frac{\Bar{\eta}^2+\Bar{\xi}^2}{4\sigma^2}+i\Delta E\,\Bar{\xi}}\frac{\theta(\Bar{\eta})}{\Bar{\eta}^2-d^2-i\,\Bar{\eta}\,\epsilon},\nonumber\\
    ~&\simeq& \frac{\sigma\sqrt{\pi}\,e^{-\Delta E^2\sigma^2}}{2\pi^2}\int_{-\infty}^{\infty}d\Bar{\eta}\,e^{-\Bar{\eta}^2/(4\sigma^2)}\, \frac{\theta(\Bar{\eta})}{(\Bar{\eta}-i\,\epsilon/2)^2-d^2}~.
\end{eqnarray}
\end{widetext}
To write the last expression, we have utilized the Gaussian integration formula $\int_{-\infty}^{\infty}d\Bar{\xi}\,e^{-\alpha\,(\Bar{\xi}+\beta)^2}=\sqrt{\pi/\alpha}$, and neglected terms $\mathcal{O}(\epsilon^2)$. To evaluate this integral one should express the factor $e^{-\Bar{\eta}^2/(4\sigma^2)} = (\sigma/\sqrt{\pi})\int_{-\infty}^{\infty}d\zeta\,e^{i\,\Bar{\eta}\,\zeta-\sigma^2\,\zeta^2}$, i.e., in terms of its Fourier transform. Then we have the integration
\begin{widetext}
\begin{eqnarray}\label{eq:Ie-M-Gaussian-p}
    \mathcal{I}_{\varepsilon}^{M} &=& \frac{\sigma^2\,e^{-\Delta E^2\sigma^2}}{2\pi^2}\, \int_{-\infty}^{\infty}d\zeta\,e^{-\sigma^2\,\zeta^2}\,\int_{-\infty}^{\infty}d\Bar{\eta}\,e^{i\,\Bar{\eta}\,\zeta}\, \frac{\theta(\Bar{\eta})}{(\Bar{\eta}-i\,\epsilon/2)^2-d^2}~.
\end{eqnarray}
\end{widetext}
The integral over $\Bar{\eta}$ has two poles at $\Bar{\eta}=\pm d+i\epsilon/2$, i.e., both are in the upper complex plane. This  integral has non-vanishing residue in the upper half complex plane, and one should then restrict the domain to $\zeta>0$. One should also note that after carrying out the integration and taking the limit $\epsilon\to 0$, the pole at $\Bar{\eta}=- d+i\epsilon/2$ will result in a factor of $\theta(- d)$ which vanishes for $d>0$. Whereas the other term will become $\theta(d)=1$. Then we are left out with 
\begin{eqnarray}\label{eq:Ie-M-Gaussian-pp}
    \mathcal{I}_{\varepsilon}^{M} &=& \frac{\sigma^2\,e^{-\Delta E^2\sigma^2}}{2\pi^2}\, \int_{0}^{\infty}d\zeta\,e^{-\sigma^2\,\zeta^2}\,e^{i\,d\,\zeta}\,\frac{2\,i\,\pi}{2\,d}~.
\end{eqnarray}
One can now evaluate this integral, see expression $2$ of the identities $(3.322)$ in page $336$ of \cite{gradshteyn2007}, which results in the analytical form as follows:
\begin{eqnarray}\label{eq:Ie-M-Gaussian-2}
    \mathcal{I}_{\varepsilon}^{M} 
    ~&=& \frac{i\sigma\,e^{-\Delta E^2\sigma^2}e^{-d^2/(4\sigma^2)}}{4\,d\sqrt{\pi}}\times\bigg[\erf\Big(\frac{id}{2\sigma}\Big)+1\bigg]~.\nonumber\\
\end{eqnarray}
We should mention that if one takes the Feynman propagator in Minkowski spacetime (see page $23$ of \cite{book:Birrell} for the expression of Feynman propagator in Minkowski spacetime) in this evaluation rather than expressing it in terms of the Wightman functions, then also one would have obtained the same result. 

Finally, with eternal switching, i.e., with $\kappa(\tau_{j})=1$, the non-local entangling term in the flat space looks like
\begin{eqnarray}\label{eq:Ie-M-Infinite-1}
    \mathcal{I}_{\varepsilon}^{M} 
    &=& \frac{1}{4\pi^2}\int_{-\infty}^{\infty}d\Bar{\xi}\,e^{i\Delta E\,\Bar{\xi}}\int_{0}^{\infty}d\Bar{\eta}\,\frac{1}{\Bar{\eta}^2-d^2-i\,\Bar{\eta}\,\epsilon}~.
\end{eqnarray}
In the above, the first integral over $\Bar{\xi}$ gives a Dirac delta distribution $\delta(\Delta E)$, which for $\Delta E>0$ gives a vanishing contribution. Therefore, with eternal switching, the non-local entangling term $\mathcal{I}_{\varepsilon}^{M}$ vanishes. This outcome reproduces the known result that in flat spacetime, with static detectors, one cannot harvest entanglement \cite{Koga:2018the}.

\subsubsection{Evaluation of $\mathcal{I}_{\varepsilon}^{GW}$}

In this part of the paper, we estimate the integral $\mathcal{I}_{\varepsilon}^{GW}$, which indicates the contribution in the non-local entangling term solely due to the presence of the GW. In this regard, as discussed previously, we have considered three types of linearized spacetimes, all of which consist of some GW burst. One of them is constructed out of the Gaussian function ($f(u)= \mathcal{A}\,e^{-u^2/\rho^2}$), one is constructed out of the sech-squared function ($f(u)=\mathcal{A}\, sech^2(u/\varrho)$), and finally, one is composed with tanh function ($f(u)=\mathcal{A}\,\{1+tanh(u/\lambda)\}$). We shall evaluate the integral $\mathcal{I}_{\varepsilon}^{GW}$ for each of the above mentioned scenarios.\vspace{0.4cm}

a. \underline{\emph{When} $f(u)= \mathcal{A}\,e^{-u^2/\rho^2}$}:-\vspace{0.15cm}

In this scenario, the expression of the part of the Wightman function of Eq. (\ref{eq:Wightman-fn-Gaussian-2}) generated purely due to the GW becomes
\begin{widetext}
\begin{eqnarray}\label{eq:Ie-GW-Gsn-Gaussian-1}
    G_{W_{GW}}(X_{B},X_{A}) &=& -\frac{\mathcal{A}(\Delta x^2-\Delta y^2)}{8\pi^2}\,\bigg[\frac{\rho\sqrt{\pi}}{\Delta u}\Big\{\erf\Big(\frac{u_{B}}{\rho}\Big) -\erf\Big(\frac{u_{A}}{\rho}\Big)\Big\}\bigg]\times \frac{1}{(\sigma_{M}+i\,\epsilon\,\Delta u)^2}~,\nonumber\\
    ~&=& -\frac{\mathcal{A}\,d^2}{8\pi^2}\,\bigg[\frac{\rho\sqrt{\pi}}{\Bar{\eta}}\Big\{\erf\Big(\frac{\Bar{\xi}+\Bar{\eta}}{2\rho}\Big) - \erf\Big(\frac{\Bar{\xi}-\Bar{\eta}}{2\rho}\Big)\Big\}\bigg]\times \frac{1}{(\Bar{\eta}^2-d^2-i\,\epsilon\,\Bar{\eta})^2}~.\nonumber\\
\end{eqnarray}
From this expression one can get the form of $G_{W_{GW}}(X_{A},X_{B})$ by utilizing the relation $G_{W_{GW}}(X_{A},X_{B})=G_{W_{GW}}(X_{B},X_{A})^{*}$, i.e., from the conjugate of $G_{W_{GW}}(X_{B},X_{A})$. This expression, with the Gaussian switching function $\kappa(\tau_{j}) = e^{-\tau^2_{j}/(2\sigma^2)}$, enables one to write the entire $\mathcal{I}_{\varepsilon}^{GW}$ as
\begin{eqnarray}\label{eq:Ie-GW-Gsn-Gaussian-2}
    \mathcal{I}_{\varepsilon}^{GW} = \frac{\mathcal{A}\,d^2}{8\pi^2}\int_{0}^{\infty}d\Bar{\eta}\int_{-\infty}^{\infty}d\Bar{\xi}\,e^{-\frac{\Bar{\eta}^2+\Bar{\xi}^2}{4\sigma^2}+i\Delta E\,\Bar{\xi}}\bigg[\frac{\rho\sqrt{\pi}}{\Bar{\eta}}\Big\{\erf\Big(\frac{\Bar{\xi}+\Bar{\eta}}{2\rho}\Big) -\erf\Big(\frac{\Bar{\xi}-\Bar{\eta}}{2\rho}\Big)\Big\}\bigg]\times \frac{1}{(\Bar{\eta}^2-d^2-i\,\epsilon\,\Bar{\eta})^2}~.
\end{eqnarray}
Evaluating the above integral directly and analytically is troublesome due to its expression in terms of the Error functions. However, there is a way to simplify this integration significantly. In this regard, one can express the combination of the error functions as 
\begin{eqnarray}\label{eq:Ie-GW-Gsn-Gaussian-3}
    \Big\{\erf\Big(\frac{\Bar{\xi}+\Bar{\eta}}{2\rho}\Big) - \erf\Big(\frac{\Bar{\xi}-\Bar{\eta}}{2\rho}\Big)\Big\} = \int\frac{d\rho}{\sqrt{\pi } \rho ^2} \bigg[(\Bar{\xi}-\Bar{\eta})\, e^{-\frac{(\Bar{\xi}-\Bar{\eta})^2}{4 \rho ^2}}-(\Bar{\xi}+\Bar{\eta})\, e^{-\frac{(\Bar{\xi}+\Bar{\eta})^2}{4 \rho ^2}}\bigg]~.
\end{eqnarray}
If we substitute this identity in the previous expression and perform the integration over the variable $\Bar{\xi}$, then the previous integral will take the form of:
\begin{eqnarray}\label{eq:Ie-GW-Gsn-Gaussian-4}
    \mathcal{I}_{\varepsilon}^{GW} &=& -\frac{\mathcal{A}\,d^2}{8\pi^2}\int_{0}^{\infty}d\Bar{\eta}\, \frac{e^{-\Bar{\eta}^2/4\sigma^2}\,(\rho\sqrt{\pi}/\Bar{\eta})}{(\Bar{\eta}^2-d^2-i\,\epsilon\,\Bar{\eta})^2}~\nonumber\\
    ~&& \times~ \int\,d\rho\,\frac{2 \bigg[\left(\Bar{\eta}-2 i \sigma ^2 \Delta E\right) e^{\frac{2 i \Bar{\eta} \sigma ^2 \Delta E}{\rho ^2+\sigma ^2}}+\Bar{\eta}+2 i \sigma ^2 \Delta E\bigg] \exp \bigg\{-\frac{\left(\Bar{\eta}+2 i \sigma ^2 \Delta E\right) \left(\Bar{\eta} \rho ^2+2 \sigma ^2 \left(\Bar{\eta}-i \rho ^2 \Delta E\right)\right)}{4 \sigma ^2 \left(\rho ^2+\sigma ^2\right)}\bigg\}}{\sqrt{\frac{1}{\rho ^2}+\frac{1}{\sigma ^2}} \left(\rho ^2+\sigma ^2\right)}~,\nonumber\\
    ~&=& \frac{\mathcal{A}\,d^2}{8\pi^2}\int_{0}^{\infty}d\Bar{\eta}\, \frac{e^{-\Bar{\eta}^2/4\sigma^2}\,(\rho\sqrt{\pi}/\Bar{\eta})}{(\Bar{\eta}^2-d^2-i\,\epsilon\,\Bar{\eta})^2}\nonumber\\
    ~&&\times~
    \frac{2 \sqrt{\pi } \rho  \sigma ^2 \sqrt{\frac{1}{\rho ^2}+\frac{1}{\sigma ^2}} e^{-\frac{\Bar{\eta}^2+4 \sigma ^4 \Delta E^2}{4 \sigma ^2}} \bigg\{\erf\left(\frac{\Bar{\eta}+2 i \sigma ^2 \Delta E}{2 \sqrt{\rho ^2+\sigma ^2}}\right)+\erf\left(\frac{\Bar{\eta}-2 i \sigma ^2 \Delta E}{2 \sqrt{\rho ^2+\sigma ^2}}\right)\bigg\}}{\sqrt{\rho ^2+\sigma ^2}}~.
\end{eqnarray}
\end{widetext}
Thus we have reduced the double integration in $\mathcal{I}_{\varepsilon}^{GW}$ to a single integration. We could not obtain an analytic expression for this integral by integrating over $\Bar{\eta}$. However, using numerical methods, one can easily evaluate this integration. In fact, in our study, we have sought the help of numerical integration to evaluate this integration and commented on the dependence of the entanglement harvesting profile on the passing of GWs. In this regard, see Figs. \ref{fig:Ie-Gsn-vDE}, \ref{fig:Ce-Gsn-vA}, and \ref{fig:Ce-Gsn-vDE} . We have elaborated on these figures in the next subsection.

Similar to the previous cases of $\mathcal{I}_{j}$ and $\mathcal{I}_{\varepsilon}^{M}$ from Eq. (\ref{eq:Ij-Infinite-1}) and Eq. (\ref{eq:Ie-M-Infinite-1}), if one considers eternal switching, i.e., with $\kappa(\tau_{j})=1$, then the non-local entangling term due to the GW can be expressed as
\begin{widetext}
\begin{eqnarray}\label{eq:Ie-GW-Gsn-Infinite-1}
    \mathcal{I}_{\varepsilon}^{GW} = \frac{\mathcal{A}\,d^2}{8\pi^2}\int_{0}^{\infty}d\Bar{\eta}\int_{-\infty}^{\infty}d\Bar{\xi}\,e^{i\Delta E\,\Bar{\xi}}\bigg[\frac{\rho\sqrt{\pi}}{\Bar{\eta}}\Big\{\erf\Big(\frac{\Bar{\xi}+\Bar{\eta}}{2\rho}\Big) -\erf\Big(\frac{\Bar{\xi}-\Bar{\eta}}{2\rho}\Big)\Big\}\bigg]\times \frac{1}{(\Bar{\eta}^2-d^2-i\,\epsilon\,\Bar{\eta})^2}~.
\end{eqnarray}
Here also, we use the integral representation of the Error function from Eq. (\ref{eq:Ie-GW-Gsn-Gaussian-3}). Then integrating over the variable $\Bar{\xi}$, one can get the quantity $\mathcal{I}_{\varepsilon}^{GW}$ as
\begin{eqnarray}\label{eq:Ie-GW-Gsn-Infinite-2}
    \mathcal{I}_{\varepsilon}^{GW} &=& -\frac{\mathcal{A}\,d^2}{8\pi^2}\int_{0}^{\infty}d\Bar{\eta}\,\frac{(\rho\sqrt{\pi}/\Bar{\eta})}{(\Bar{\eta}^2-d^2-i\,\epsilon\,\Bar{\eta})^2}~\times~\int\,d\rho\, \Big[8 \rho\,  \Delta E\, e^{-\rho ^2\, \Delta E^2} \sin (\Bar{\eta} \Delta E)\Big],\nonumber\\
    ~&=& \frac{\mathcal{A}\,d^2}{8\pi^2}\int_{0}^{\infty}d\Bar{\eta}\, \frac{(\rho\sqrt{\pi}/\Bar{\eta})}{(\Bar{\eta}^2-d^2-i\,\epsilon\,\Bar{\eta})^2}~\times~
    \Big[ \frac{4 e^{-\rho ^2 \,\Delta E^2} \sin (\Bar{\eta} \Delta E)}{\Delta E}\Big]~.
\end{eqnarray}
\end{widetext}
Now, for this integral also, we provide a numerical result. It should be noted that with eternal switching, i.e., when the detectors interact with the background field for an infinite time, the individual detector transition vanishes, see Eq. (\ref{eq:Ij-Infinite-1}) and the discussions thereafter. We have also observed in Eq. (\ref{eq:Ie-M-Infinite-1}) that the contribution from non-local entangling terms of flat space $\mathcal{I}_{\varepsilon}^{M}$ vanishes. Therefore, in this case, the entire entanglement harvesting measure, i.e., concurrence, is dictated by the $\mathcal{I}_{\varepsilon}^{GW}$ term. We have plotted this quantity in Fig. \ref{fig:Ie-Inf-Gsn-sech2-vDE}, and the discussion on this figure is given in our next subsection. {  Moreover, we find that as $\Delta E\to 0$, the integral reaches a finite value. We will see later how the behaviour of this integral will signify the presence and absence of memory as the detector energy gap is lowered.}\vspace{0.4cm}

b. \underline{\emph{When} $f(u)=\mathcal{A}\,sech^2(u/\varrho)$} :-\vspace{0.15cm}

Let us now check the consequences if the GW burst is of the form $f(u)=\mathcal{A}\,sech^2(u/\varrho)$. As we have seen previously, for eternal switching, i.e., for $\kappa(\tau_{j})=1$, the integrals $\mathcal{I}_{j}=0=\mathcal{I}_{\varepsilon}^{M}$. Then the entire concurrence becomes only dependent on the integral $\mathcal{I}_{\varepsilon}^{GW}$, in fact in that scenario the concurrence is given by $\mathcal{C}_{\mathcal{I}} = |\mathcal{I}_{\varepsilon}^{GW}|$. Also with non-trivial finite time switching the estimation of $\mathcal{I}_{ \varepsilon}^{GW}$ becomes highly complicated and often cannot be pursued analytically. Therefore, this situation of $\kappa(\tau_{j})=1$ becomes more suitable for investigating various effects of different GW profiles in the harvested entanglement. Then, for this certain GW burst, let us consider the eternal switching and evaluate the relevant integral. In this case, with the help of Green's function Eq. (\ref{eq:Wightman-fn-sech2-1}), the expression for $\mathcal{I}_{\varepsilon}^{GW}$ takes the initial form
\begin{widetext}
\begin{eqnarray}\label{eq:Ie-GW-Sec2-Infinite-1}
    \mathcal{I}_{\varepsilon}^{GW} &=& \frac{\mathcal{A}\,d^2}{4\pi^2}\int_{0}^{\infty}d\Bar{\eta}\int_{-\infty}^{\infty}d\Bar{\xi}\,e^{i\Delta E\,\Bar{\xi}}\bigg[\frac{\varrho}{\Bar{\eta}}\Big\{\tanh{\Big(\frac{\Bar{\xi}+\Bar{\eta}}{2\varrho}\Big)} -\tanh{\Big(\frac{\Bar{\xi}-\Bar{\eta}}{2\varrho}\Big)}\Big\}\bigg]\times \frac{1}{(\Bar{\eta}^2-d^2-i\,\epsilon\,\Bar{\eta})^2}~\nonumber\\
    &=& \frac{\mathcal{A}\,d^2}{4\pi^2}\int_{0}^{\infty} \frac{d\Bar{\eta}}{(\Bar{\eta}^2-d^2-i\,\epsilon\,\Bar{\eta})^2}\,\frac{\varrho}{\Bar{\eta}}\times
    \int_{-\infty}^{\infty}d\Bar{\xi}\,e^{i\Delta E\,\Bar{\xi}}\bigg[\Big\{\tanh{\Big(\frac{\Bar{\xi}+\Bar{\eta}}{2\varrho}\Big)} -\tanh{\Big(\frac{\Bar{\xi}-\Bar{\eta}}{2\varrho}\Big)}\Big\}\bigg]~,
\end{eqnarray}
where we assume the eternal switching condition. Now one can express $\tanh{x}$ as a sum, see expansion $2$ of the identities $(1.421)$ in page $44$ of \cite{gradshteyn2007}, as 
\begin{eqnarray}\label{eq:Ie-GW-Sec2-Infinite-2}
    \tanh{x} = \sum_{k=1}^{\infty}\bigg[\frac{1}{x+i \pi  \left(k-\frac{1}{2}\right)}+\frac{1}{x-i \pi  \left(k-\frac{1}{2}\right)}\bigg]~. 
\end{eqnarray}
Using this expansion, one can observe that 
\begin{eqnarray}\label{eq:Ie-GW-Sec2-Infinite-3}
    \tanh{\Big(\frac{\Bar{\xi}+\Bar{\eta}}{2\varrho}\Big)} -\tanh{\Big(\frac{\Bar{\xi}-\Bar{\eta}}{2\varrho}\Big)} &=& \sum_{k=1}^{\infty}\bigg[\frac{1}{(\Bar{\xi}+\Bar{\eta})/(2 \varrho )-i \pi  \left(k-\frac{1}{2}\right)}+\frac{1}{(\Bar{\xi}+\Bar{\eta})/(2 \varrho )+i \pi  \left(k-\frac{1}{2}\right)}\nonumber\\
    ~&~&~~~~~~ -\frac{1}{(\Bar{\xi}-\Bar{\eta})/(2 \varrho )-i \pi  \left(k-\frac{1}{2}\right)}-\frac{1}{(\Bar{\xi}-\Bar{\eta})/(2 \varrho )+i \pi  \left(k-\frac{1}{2}\right)}\bigg]~. 
\end{eqnarray}
\end{widetext}
Now one can notice that the first and the third quantity in the sum of the previous equation have poles of order unity in the upper half complex plane. At the same time, the second and the fourth quantity have poles in the lower half complex plane. Furthermore, for $\Delta E>0$, one has to consider a contour in the complex upper plane to dampen the relevant integrals to evaluate the integral of Eq. (\ref{eq:Ie-GW-Sec2-Infinite-1}). Therefore, in the result of that integral, only the first and the third quantity of Eq. (\ref{eq:Ie-GW-Sec2-Infinite-3}) will give a non-zero contribution. Taking these contributions properly and summing over the integer $k$, one can get the expression of $\mathcal{I}_{\varepsilon}^{GW}$ as
\begin{eqnarray}\label{eq:Ie-GW-Sec2-Infinite-4}
    \mathcal{I}_{\varepsilon}^{GW} &=& \frac{\mathcal{A}\,d^2}{4\pi^2}\int_{0}^{\infty}d\Bar{\eta}\,\frac{(\varrho/\Bar{\eta})}{(\Bar{\eta}^2-d^2-i\,\epsilon\,\Bar{\eta})^2}\Big[\frac{4\pi\varrho\,\sin (\Bar{\eta} \Delta E)}{\sinh{(\pi\Delta E\,\varrho)}}\Big]~.
\end{eqnarray}
Thus the quantity $\mathcal{I}_{\varepsilon}^{GW}$ has been reduced to a single integration form, and we have employed numerical methods to obtain the final result. Comparing Eq. (\ref{eq:Ie-GW-Sec2-Infinite-4}) and the previous Eq. (\ref{eq:Ie-GW-Gsn-Infinite-2}) one can notice that the $\Bar{\eta}$ integration in both the places are done over the same function. Therefore, their characteristics should also be the same. However, they have different multiplicative factors depending on other system parameters. It is to see whether the overall characteristics of $|\mathcal{I}_{\varepsilon}^{GW}|$ remain the same in both cases. {  As mentioned previously for the Gaussian case, likewise we find that as $\Delta E \to 0$, the integral (\ref{eq:Ie-GW-Sec2-Infinite-4}) remains finite.} The corresponding plots are given in Fig. \ref{fig:Ie-Inf-Gsn-sech2-vDE}.\vspace{0.4cm}

c. \underline{\emph{When} $f(u)=\mathcal{A}\,\theta(u)$} :-\vspace{0.15cm}

Now we consider GW profile of the form $f(u)=\mathcal{A}\,\theta(u)$ with eternal switching $\kappa(\tau_{j})=1$. As mentioned previously with this eternal switching $\mathcal{I}_{j}=0=\mathcal{I}_{\varepsilon}^{M}$ and the entire concurrence is given by the quantity $\mathcal{C}_{\mathcal{I}}=|\mathcal{I}_{\varepsilon}^{GW}|$. Let us now evaluate this $\mathcal{I}_{\varepsilon}^{GW}$ that arrives solely due to the presence of the gravitational wave burst. In this regard, with the help of Eq. (\ref{eq:Wightman-fn-thetaFn-1}) we get
\begin{widetext}
\begin{eqnarray}\label{eq:Ie-GW-thetaFn-Infinite-1}
    \mathcal{I}_{\varepsilon}^{GW} &=& \frac{\mathcal{A}\,d^2}{4\pi^2}\int_{0}^{\infty}\frac{d\Bar{\eta}}{\Bar{\eta}}\frac{1}{(\Bar{\eta}^2-d^2-i\,\epsilon\,\Bar{\eta})^2} \int_{-\infty}^{\infty}d\Bar{\xi}\,e^{i\Delta E\,\Bar{\xi}}\bigg[\Big(\frac{\Bar{\xi}+\Bar{\eta}}{2}\Big)\,\theta\Big(\frac{\Bar{\xi}+\Bar{\eta}}{2}\Big) -\Big(\frac{\Bar{\xi}-\Bar{\eta}}{2}\Big)\,\theta\Big(\frac{\Bar{\xi}-\Bar{\eta}}{2}\Big)\bigg],~\nonumber\\
    &=& \frac{i\,\mathcal{A}\,d^2}{4\pi^2\Delta E^2}\int_{0}^{\infty}d\Bar{\eta}\,\frac{\sin (\Bar{\eta} \Delta E)}{\Bar{\eta}}\,\frac{1}{(\Bar{\eta}^2-d^2-i\,\epsilon\,\Bar{\eta})^2}~.
\end{eqnarray}
\end{widetext}
One can pursue this integral numerically. We plot the modulus of this quantity, i.e., the concurrence, in Fig. \ref{fig:Ie-Inf-thetaFn-vDE} and compare its differences and resemblances with other scenarios. Note a crucial difference between  Eq. (\ref{eq:Ie-GW-thetaFn-Infinite-1}) and Eqs. (\ref{eq:Ie-GW-Sec2-Infinite-2}), (\ref{eq:Ie-GW-Sec2-Infinite-4}) in the limit $\Delta E\to 0$. These integrals are for bursts with and without memory, respectively. In the latter one, we find that as the detector transition energy level goes to zero, the integral settles to a constant. But in the former one, we find a $\mathcal{O}(1/\Delta E)$ term. This is akin to the Weinberg term in the soft theorems, which scales as $\mathcal{O}(1/\omega)$  \cite{Strominger:2017}. The memory term is a simple Fourier transform of the soft graviton factor \cite{Strominger:2017}. In our calculation also, for the {\em step-function} memory profile we recover such term because $\mathcal{I}_{\varepsilon}^{GW}(\Delta E)$ is nearly a Fourier transform of the Wightman function which is linear in the burst profile. Since the universal feature of a gravitational wave burst is almost like a step-function \cite{Favata:2010}, we get that for such profiles the concurrence measure  does not reach a finite value, as the detector energy gap is reduced, unlike the burst profiles without the memory effect. Thus, this is the crucial difference that determines the nature of entanglement harvesting between profiles with and without memory.\vspace{0.4cm} 
\color{black}

d. \underline{\emph{When} $f(u)=\mathcal{A}\,\{1+tanh(u/\lambda)\}$} :-\vspace{0.15cm}

In this part, we consider a GW profile of the form $f(u)=\mathcal{A}\,\{1+tanh(u/\lambda)\}$, which will retain GW memory to an asymptotic future observer. Here also, like the previous case, we consider the eternal switching, i.e., $\kappa(\tau_{j})=1$. One of the reasons is, in this scenario $\mathcal{I}_{j}=0=\mathcal{I}_{\varepsilon}^{M}$ and thus the concurrence becomes entirely expressed by a single integral $\mathcal{C}_{\mathcal{I}}=|\mathcal{I}_{\varepsilon}^{GW}|$. Therefore, all quantities in concurrence that arise due to some non-trivial switching are zero in this case, and one can focus on the quantities that are solely due to the background. Moreover, with this particular switching, the evaluation of the integral $\mathcal{I}_{\varepsilon}^{GW}$, in this case, can be analytically followed up to a certain point. With these considerations, let us now evaluate the integral $\mathcal{I}_{\varepsilon}^{GW}$ that can be expressed, with the help of Eq. (\ref{eq:Wightman-fn-tanh-1}), as
\begin{widetext}
\begin{eqnarray}\label{eq:Ie-GW-tanh-Infinite-1}
    \mathcal{I}_{\varepsilon}^{GW} &=& \frac{\mathcal{A}\,d^2}{4\pi^2}\int_{0}^{\infty}d\Bar{\eta}\int_{-\infty}^{\infty}d\Bar{\xi}\,e^{i\Delta E\,\Bar{\xi}}\bigg[1+\frac{\lambda}{\Bar{\eta}}\Big\{\ln\Big[\cosh{\Big(\frac{\Bar{\xi}+\Bar{\eta}}{2\lambda}\Big)}\Big] -\ln\Big[\cosh{\Big(\frac{\Bar{\xi}-\Bar{\eta}}{2\lambda}\Big)}\Big]\Big\}\bigg]\times \frac{1}{(\Bar{\eta}^2-d^2-i\,\epsilon\,\Bar{\eta})^2},~\nonumber\\
    &=& \frac{\mathcal{A}\,d^2}{4\pi^2}\int_{0}^{\infty} \frac{d\Bar{\eta}}{(\Bar{\eta}^2-d^2-i\,\epsilon\,\Bar{\eta})^2}\,\frac{\lambda}{\Bar{\eta}}\times
    \int_{-\infty}^{\infty}d\Bar{\xi}\,e^{i\Delta E\,\Bar{\xi}}\bigg[\ln\Big[\cosh{\Big(\frac{\Bar{\xi}+\Bar{\eta}}{2\lambda}\Big)}\Big] -\ln\Big[\cosh{\Big(\frac{\Bar{\xi}-\Bar{\eta}}{2\lambda}\Big)}\Big]\bigg]~.
\end{eqnarray}
Now one can express the logarithmic quantities inside the bracket in the previous expression as
\begin{eqnarray}\label{eq:Ie-GW-tanh-Infinite-2}
    \ln\Big[\cosh{\Big(\frac{\Bar{\xi}+\Bar{\eta}}{2\lambda}\Big)}\Big] -\ln\Big[\cosh{\Big(\frac{\Bar{\xi}-\Bar{\eta}}{2\lambda}\Big)}\Big] = -\int\,\frac{d\lambda}{2 \lambda ^2}\,\Big[(\Bar{\xi}+\Bar{\eta}) \tanh \left(\frac{\Bar{\xi}+\Bar{\eta}}{2 \lambda }\right)-(\Bar{\xi}-\Bar{\eta}) \tanh \left(\frac{\Bar{\xi}-\Bar{\eta}}{2 \lambda }\right)\Big]~.
\end{eqnarray}
\end{widetext}
Furthermore, one can express these $\tanh$ functions as a sum provided in Eq. (\ref{eq:Ie-GW-Sec2-Infinite-2}), which enables one to understand the pole structure of the integrand. The expansion will be the same as given in Eq. (\ref{eq:Ie-GW-Sec2-Infinite-3}), with the first two quantities in the sum multiplied by $(\Bar{\xi}+\Bar{\eta})$ and the last two terms multiplied by $(\Bar{\xi}-\Bar{\eta})$, and the $\varrho$ replaced by $\lambda$. Then we carry out the $\Bar{\xi}$ integral first. It is to be noted that there is a factor of $e^{i\Delta E\,\Bar{\xi}}$ in the integral, where by our choice $\Delta E>0$. Therefore, to carry out this integration one must choose a contour in the upper half complex plane. The terms analogous to the first and third quantities in Eq. (\ref{eq:Ie-GW-Sec2-Infinite-3}) will contain poles in the upper half complex plane, and only these terms will contribute to the integration. After this integration, one can also easily perform the sum over $k$ and the integration over $\lambda$. Then the expression of $\mathcal{I}_{\varepsilon}^{GW}$ only has an integration over $\Bar{\eta}$, and this expression looks like
\begin{eqnarray}\label{eq:Ie-GW-tanh-Infinite-4}
    \mathcal{I}_{\varepsilon}^{GW} &=& \frac{\mathcal{A}\,d^2}{4\pi^2}\int_{0}^{\infty}d\Bar{\eta}\,\frac{(\lambda/\Bar{\eta})}{(\Bar{\eta}^2-d^2-i\,\epsilon\,\Bar{\eta})^2}\Big[\frac{2\pi\,\cos (\Bar{\eta} \Delta E)}{\Delta E\,\sinh{(\pi\,\Delta E\,\lambda)}}\Big]~.
\end{eqnarray}
This integral can be carried out numerically and it will be interesting to compare the outcome from it with Eqs. (\ref{eq:Ie-GW-Gsn-Infinite-2}) and (\ref{eq:Ie-GW-Sec2-Infinite-4}). We plot the modulus of this quantity in Fig. \ref{fig:Ie-Inf-tanh-vDE} and discuss the consequences in the next subsection. {  Here we find that, unlike bursts without memory, the integral falls as $\mathcal{O}(1/\Delta E^2)$ as $\Delta E\to 0$. }

\subsection{Estimation of the concurrence}\label{subsec:concurrence}

%%%%%%%%%%%%%%%%%%%%%%%%%%%%%%%%%%%%%%%%%%%%%%%%%%%%%%%%%%%%%%%%%%%%%%%%%%%%%%%%%%%%%%%

Let us now discuss the measure of harvested entanglement, and understand how different burst profiles shape this quantity. We are also interested to understand what contrasting features these bursts may bring to the fore and compare the results with  the periodic waveform discussed earlier in literature \cite{Xu:2020pbj}. In passing, we also aim to highlight the differences in entanglement harvesting, if any, between the two different types of burst profiles, i.e., with or without memory. {  Since the Wightman function appears in Concurrence expression linearly, thus, we have only $\mathcal{O}(\mathcal{A})$ terms in Concurrence. No higher-order terms appear that need to be dropped in our calculation. We have discussed more on the consistency of this perturbative approach in Appendix \ref{Appn:perturbation-consistency}.}

\subsubsection{When $f(u)= \mathcal{A}\,e^{-u^2/\rho^2}$}

\begin{figure*}[htp]
\centering
\includegraphics[width=0.47\linewidth]{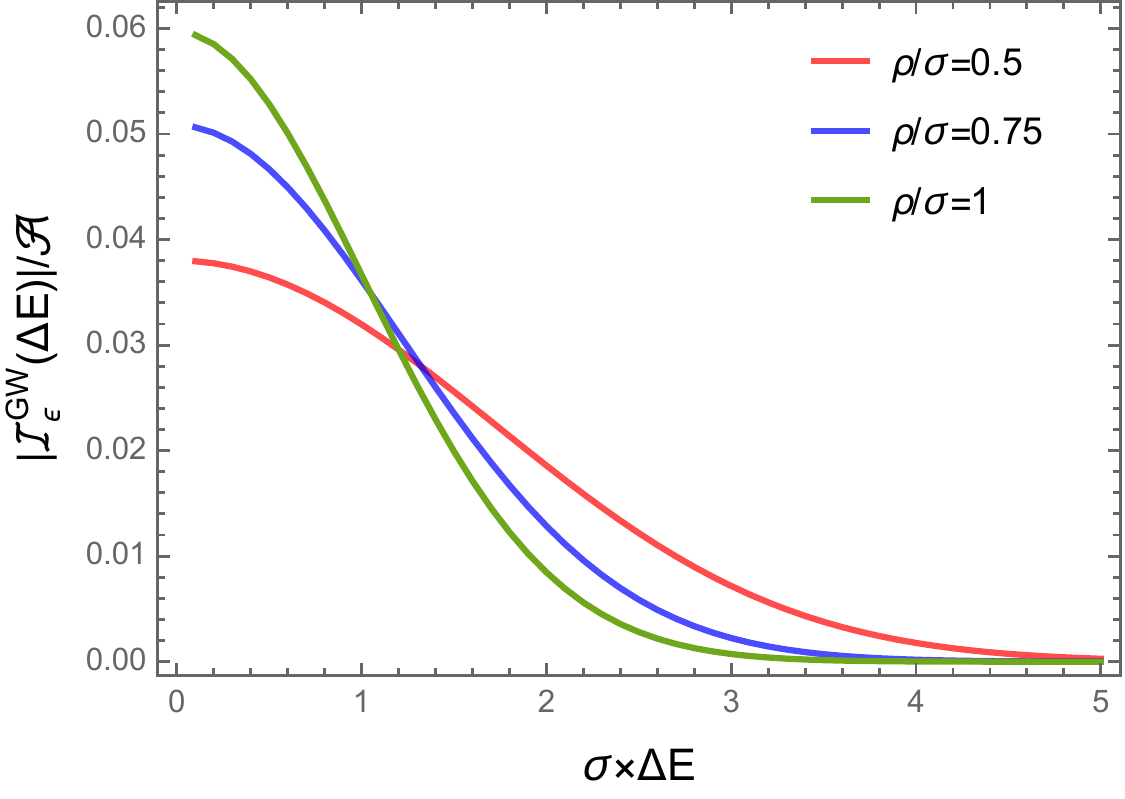}
\hskip 10pt
\includegraphics[width=0.47\linewidth]{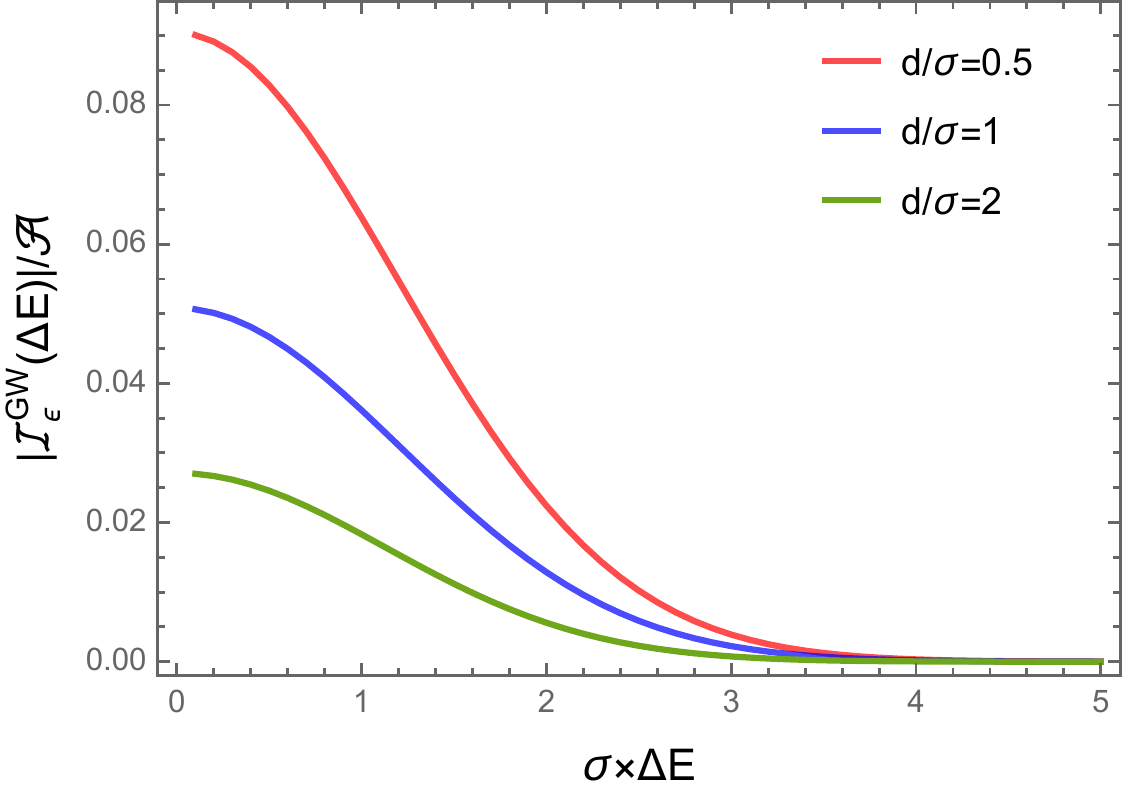}
\caption{The modulus of the quantity $\mathcal{I}_{\varepsilon}^{GW}(\Delta E)$ is plotted as a function of the dimensionless detector transition energy $(\sigma\,\Delta E)$. This quantity corresponds to the contribution of a GW with $f(u)=\mathcal{A}\, e^{-u^2/\rho^2}$ in entanglement harvesting. Both of the above plots correspond to detectors that interact with the background field through Gaussian window functions, i.e., the switching is $\kappa(\tau_{j})=e^{-\tau^2_{j}/(2\sigma^2)}$. In the left plot, different curves correspond to different $\rho/\sigma$, and the other parameters are fixed at $d/\sigma=1$. On the other hand, in the right plot, different curves correspond to different distances $d/\sigma$ between the static detectors, and $\rho/\sigma=0.75$ is fixed. These plots show that when detector transition energy is low, one can get higher harvesting for higher $\rho/\sigma$. Whereas, for moderately large transition energy, one gets more harvesting for lower $\rho/\sigma$. On the other hand, entanglement harvesting is always larger for a smaller distance between the two detectors. Furthermore, most importantly, harvesting decreases with increasing transition energy.}\label{fig:Ie-Gsn-vDE}
\end{figure*}

\begin{figure}[htp]
\centering
\includegraphics[width=8.50cm]{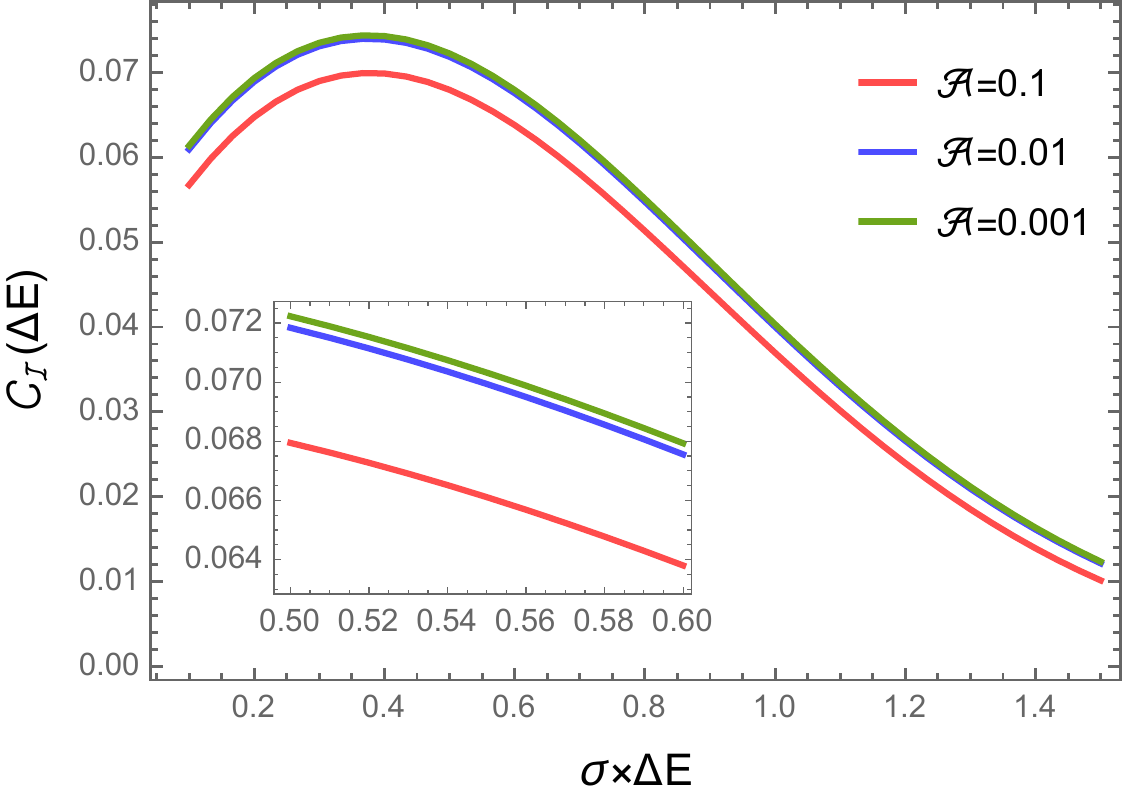}
\caption{The concurrence $\mathcal{C}_{\mathcal{I}}(\Delta E)$ is plotted as a function of the dimensionless detector transition energy $(\sigma\,\Delta E)$ for different perturbation strength $\mathcal{A}$ of the GW ($f(u)=\mathcal{A}\, e^{-u^2/\rho^2}$). The above plot corresponds to detectors that interact with the background field through Gaussian window functions $\kappa(\tau_{j})=e^{-\tau^2_{j}/(2\sigma^2)}$. The other parameters are fixed at $\rho/\sigma=0.75$ and $d/\sigma=1$. It is observed that the GW diminishes entanglement harvesting, as a greater perturbation strength results in lesser entanglement harvesting.}\label{fig:Ce-Gsn-vA}
\end{figure}

\begin{figure*}[htp]
\centering
\includegraphics[width=0.47\linewidth]{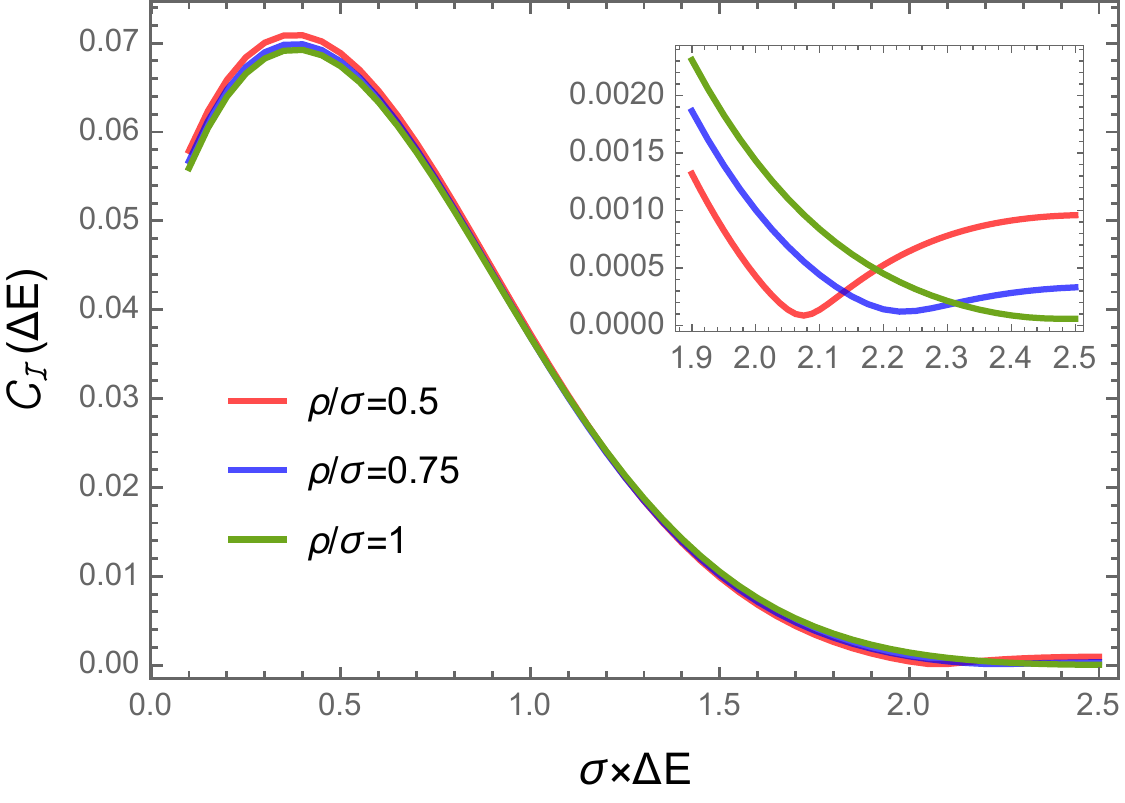}
\hskip 10pt
\includegraphics[width=0.47\linewidth]{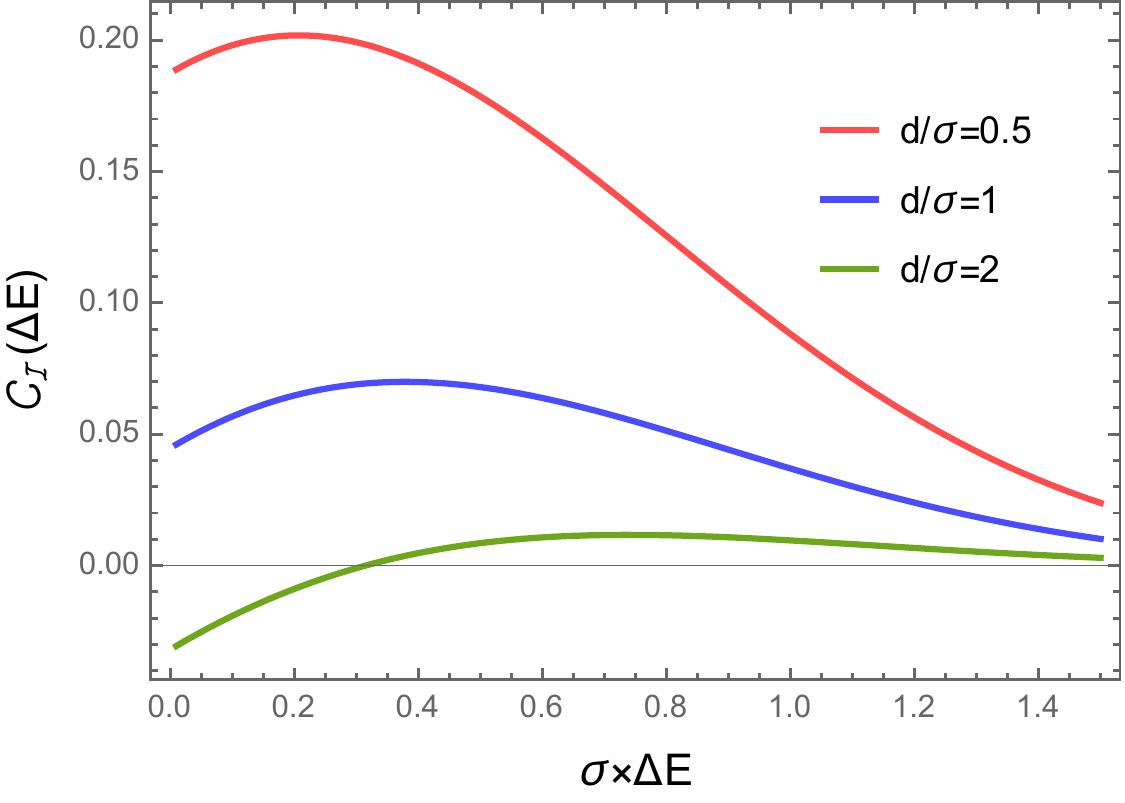}
\caption{The concurrence $\mathcal{C}_{\mathcal{I}}(\Delta E)$ is plotted as a function of the dimensionless detector transition energy $(\sigma\,\Delta E)$. Both of the above plots correspond to detectors that interact with the background field through Gaussian window functions $\kappa(\tau_{j})=e^{-\tau^2_{j}/(2\sigma^2)}$ with $f(u)=\mathcal{A}\, e^{-u^2/\rho^2}$. In the left plot, we have fixed $\rho/\sigma=0.75$; in the right plot, we have fixed $d/\sigma=1$. In both of the above plots, we have considered $\mathcal{A}=0.1$.
From the left plots, we observe that the entanglement harvesting is larger for smaller $\rho/\sigma$ in low $(\sigma\,\Delta E\lesssim 1.1)$ and high $(\sigma\,\Delta E\gtrsim 2.2)$ detector transition energy regimes. However, harvesting increases with increasing $\rho/\sigma$ in the intermediate transition energy regime. In this regard, the epilogue of the left figure provides a clear picture.
The right plots confirm that entanglement harvesting decreases with increasing distance $d/\sigma$. For large distances, like when $d/\sigma=2$, entanglement harvesting begins after a certain minimum $(\sigma\,\Delta E)$. However, later we shall see that this last phenomenon is not generic and is switching-dependent, as it is not present for eternal switching.}\label{fig:Ce-Gsn-vDE}
\end{figure*}

\begin{figure*}
\centering
\includegraphics[width=0.47\linewidth]{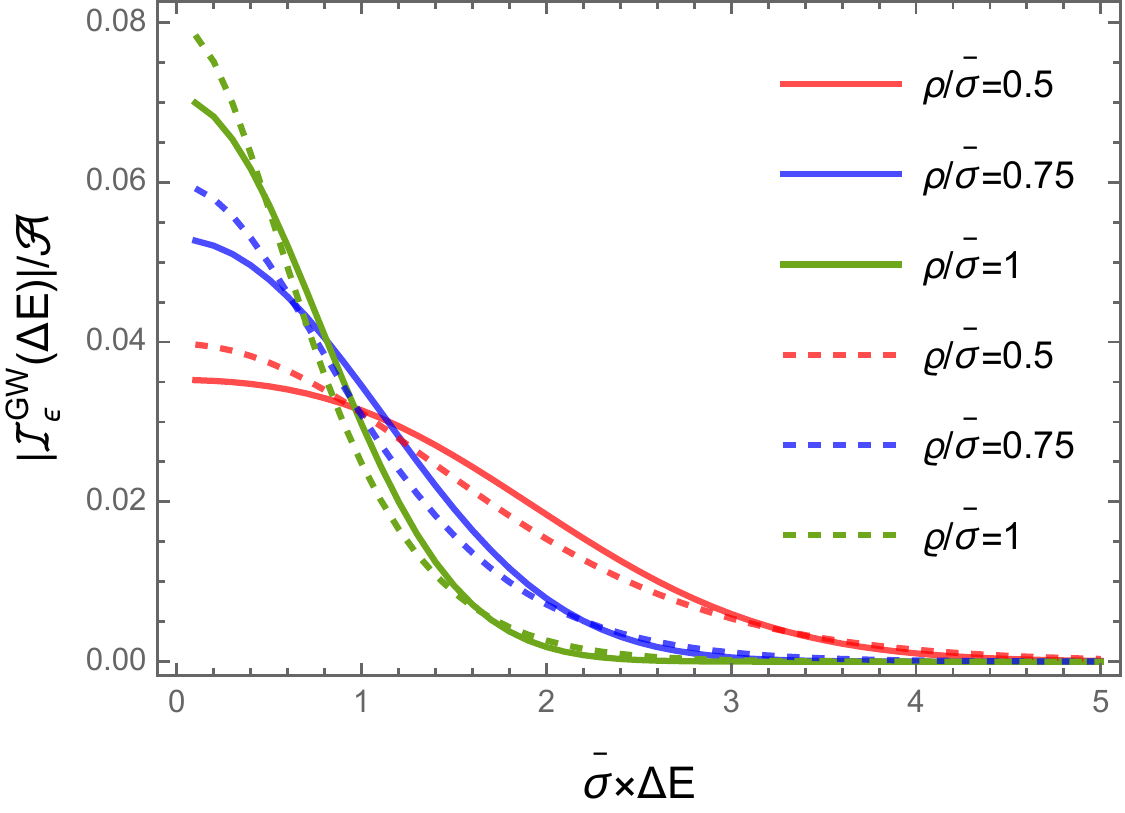}
\hskip 10pt
\includegraphics[width=0.47\linewidth]{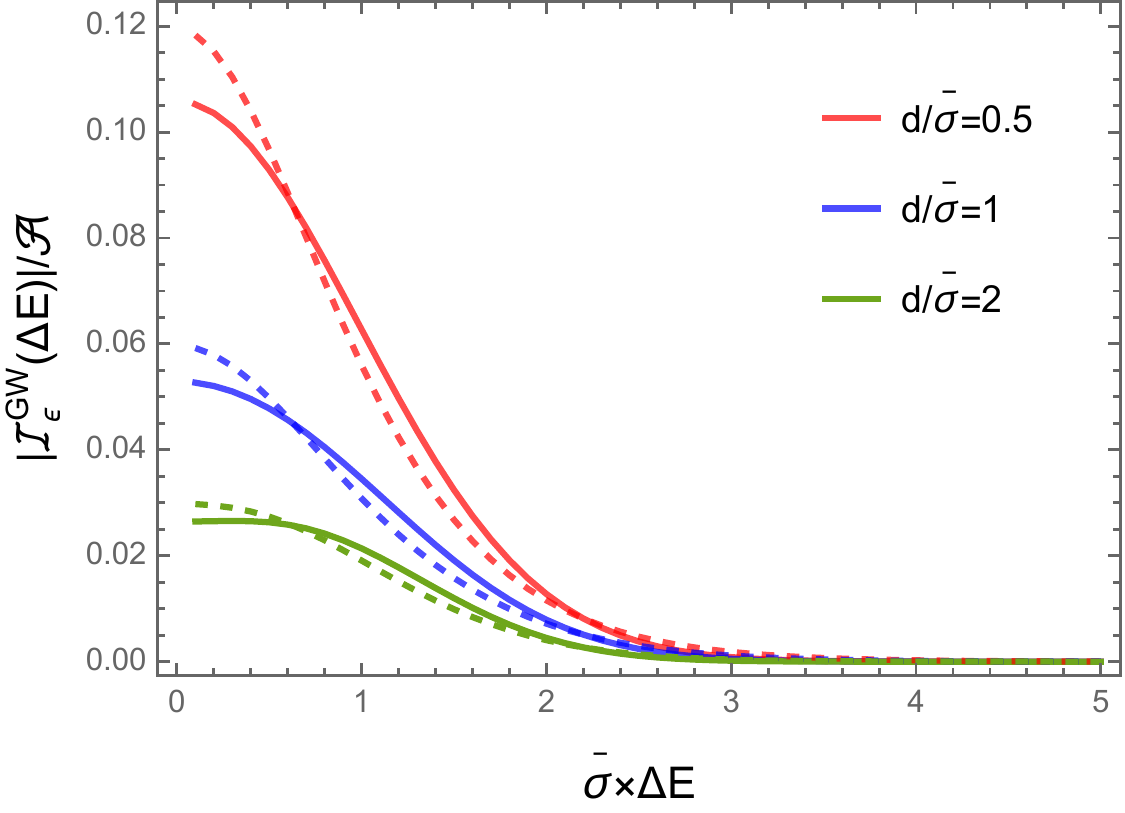}
\caption{The modulus of the quantity $\mathcal{I}_{\varepsilon}^{GW}(\Delta E)$ is plotted as a function of the dimensionless detector transition energy $(\bar{\sigma}\,\Delta E)$ for $f(u)=\mathcal{A}\, e^{-u^2/\rho^2}$ (solid curves) and $f(u)=\mathcal{A}\,sech^2(u/\varrho)$ (dashed curves). Both of the above plots correspond to detectors interacting with the background field for infinite time, i.e., with switching function $\kappa(\tau_{j})=1$. Therefore, the quantity $|\mathcal{I}_{\varepsilon}^{GW}(\Delta E)|$ is the same as concurrence. Note, here we have introduced a new dimension-full parameter $\bar{\sigma}$ to make the other parameters dimensionless, which makes the comparison with the Gaussian switching case much easier. In the left plot, different curves correspond to different $\rho/\Bar{\sigma}$ (or $\varrho/\Bar{\sigma}$), while $d/\Bar{\sigma}=1$ is fixed. In the right plot, different curves correspond to different distances $d/\Bar{\sigma}$ between the static detectors, and we have fixed $\rho/\Bar{\sigma}=0.75=\varrho/\Bar{\sigma}$. The qualitative features of these plots are the same as that of the Gaussian GW burst with the Gaussian switching. Here also, we observe that when detector transition energy is low, one can get higher harvesting for higher $\rho/\Bar{\sigma}$ (or $\varrho/\Bar{\sigma}$), for moderately large transition energy, one gets more harvesting for lower $\rho/\Bar{\sigma}$ (or $\varrho/\Bar{\sigma}$). Furthermore, entanglement harvesting is always larger for a smaller distance between the two detectors, and harvesting decreases with increasing transition energy. It should be noted that in these figures, the quantitative values of the concerned quantity are slightly higher than that of Fig. \ref{fig:Ie-Gsn-vDE}.}\label{fig:Ie-Inf-Gsn-sech2-vDE}
\end{figure*}

In a manner similar to our previous discussions, here also, we first consider the GW burst with $f(u)= \mathcal{A}\,e^{-u^2/\rho^2}$. We take the expressions of $\mathcal{I}_{j}$, $\mathcal{I}_{ \varepsilon}^{M}$, and  $\mathcal{I}_{\varepsilon}^{GW}$ from Eqs. (\ref{eq:Ij-Gaussian-3}), (\ref{eq:Ie-M-Gaussian-2}), and (\ref{eq:Ie-GW-Gsn-Gaussian-4}) respectively, and have plotted $|\mathcal{I}_{\varepsilon}^{GW}(\Delta E)|$ as a function of the dimensionless detector transition energy $(\sigma\,\Delta E)$ in Fig. \ref{fig:Ie-Gsn-vDE}. Naturally, this quantity in entanglement harvesting corresponds to the contribution of the GWs only. These plots relate to the Gaussian window functions $\kappa(\tau_{j})=e^{-\tau^2_{j}/(2\sigma^2)}$ as evident from the consideration of Eq. (\ref{eq:Ie-GW-Gsn-Gaussian-4}). From these plots, one can conclude the following remarks:
%%%
\begin{itemize}
    \item The effect of GWs on entanglement harvesting decreases with increasing detector transition energy.

    \item For low transition energy of the detector, one can get higher harvesting for higher $\rho/\sigma$.

    \item With moderately large transition energy, one gets more harvesting for lower $\rho/\sigma$.

    \item The entanglement harvesting is always larger for shorter distances between the two detectors.
\end{itemize}
%%%
%
%%%%%%%%
%%%%%%%%%%%%%%%%
%%%%%%%%%%%%%%%%%%%%%%%%%%%%%%%%%%%%%%%%%%%%%%%%%%%%%%%%%%%%%%%%%
%%%%%%%%%%%%%%%%%%%%%%%%%%%%%%%%%%%%%%%%%%%%%%%%%%%%%%%%%%%%%%%
In Fig. \ref{fig:Ce-Gsn-vA} and \ref{fig:Ce-Gsn-vDE}, we have plotted the entire concurrence as a function of the dimensionless transition energy $(\sigma\,\Delta E)$, with  $f(u)= \mathcal{A}\,e^{-u^2/\rho^2}$ and the Gaussian switching $\kappa(\tau_{j})=e^{-\tau^2_{j}/(2\sigma^2)}$. In particular, Fig. \ref{fig:Ce-Gsn-vA} depicts plots for varying $\mathcal{A}$, whereas, Fig. \ref{fig:Ce-Gsn-vDE} depicts plots for varying $\rho/\sigma$ and $d/\sigma$. The key features of these plots can be summarised below.
%%%
\begin{itemize}
    \item Fig. \ref{fig:Ce-Gsn-vA} suggests that with decreasing $\mathcal{A}$ the harvesting increases. It indicates that, with the Gaussian switching, in the non-local entangling term $\mathcal{I}_{\varepsilon}=\mathcal{I}_{\varepsilon}^{M} + \mathcal{I}_{\varepsilon}^{GW}$, the individual contributions of $\mathcal{I}_{\varepsilon}^{M}$ and $\mathcal{I}_{\varepsilon}^{GW}$ oppose each other.

    \item  {  Note in Fig. \ref{fig:Ce-Gsn-vA} that as one decreases the value of $\mathcal{A}$, the concurrence asymptotes to a final value.}  

    \item From the left plot of Fig. \ref{fig:Ce-Gsn-vDE}, we observe that in low $(\sigma\,\Delta E\lesssim 1.1)$ and high $(\sigma\,\Delta E\gtrsim 2.2)$ detector transition energy regimes, the entanglement harvesting is larger for smaller $\rho/\sigma$. At the same time, harvesting is larger for larger $\rho/\sigma$ in the intermediate transition energy regime. Therefore, in low and intermediate transition energy regimes, the characteristics of the concurrence and $|\mathcal{I}_{\varepsilon}^{GW}(\Delta E)|$, as seen from Fig. \ref{fig:Ie-Gsn-vDE} and Fig. \ref{fig:Ce-Gsn-vDE}, is opposite with respect to $\rho/\sigma$.

    \item From the right plots of Fig. \ref{fig:Ce-Gsn-vDE}, we perceive that entanglement harvesting decreases with increasing distance $d/\sigma$. Interestingly, for large distances, like when $d/\sigma=2$, entanglement harvesting begins after a certain minimum $(\sigma\,\Delta E)$. It is to be noted that this case is specific to the Gaussian switching. We also mention that the choice of $\mathcal{A}$ for depicting the concerned plots, is motivated by the need for obtaining visually distinguishable curves. Practically, the value of $\mathcal{A}$ should be much smaller, e.g., this amplitude is of the order of $\sim 10^{-21}$ for the GWs detected on earth \cite{Abbott:2016blz, Xu:2020pbj}.

    \item In Fig. \ref{fig:Ie-Inf-Gsn-sech2-vDE} we have plotted $|\mathcal{I}_{\varepsilon}^{GW}(\Delta E)|$ as function of the dimensionless energy $(\bar{\sigma}\,\Delta E)$ for $f(u)= \mathcal{A}\,e^{-u^2/\rho^2}$ and eternal switching $(\kappa(\tau_{j})=1)$. Note, here we have introduced an additional dimension-full parameter $\bar{\sigma}$ to obtain the other parameters and quantities in a dimensionless fashion. Introducing this parameter rather than choosing existing parameters to define dimensionless quantities provides a form similar to the finite interaction (with the Gaussian switching) case and makes the comparison much easier. In this scenario as $\mathcal{I}_{j}$ and $\mathcal{I}_{ \varepsilon}^{M}$ vanish (see Eq. (\ref{eq:Ij-Infinite-1}) and (\ref{eq:Ie-M-Infinite-1}) and the related discussions), the entire concurrence is actually given by the quantity $|\mathcal{I}_{\varepsilon}^{GW}(\Delta E)|$. Therefore, in Fig. \ref{fig:Ie-Inf-Gsn-sech2-vDE}, we have effectively plotted the concurrence with eternal switching. The qualitative behavior of the curves from this figure is the same as that of Fig. \ref{fig:Ie-Gsn-vDE}. However, the quantity depicted in Fig. \ref{fig:Ie-Inf-Gsn-sech2-vDE} has a bit larger value compared to \ref{fig:Ie-Gsn-vDE}. This figure also signifies that the GW burst of our considered form can always induce entanglement, even between static detectors. Obviously, the eternal interaction scenario makes this claim more prominent in this case, as $\mathcal{I}_{j}$ vanishes.
\end{itemize}
%%%

%%%%%%%%%%%%%%%%%%%%%%%%%%%%%%%%%%%%%%%%%%%%%%%%%%%%%%%%%%%%%%%%%

%%%%%%%%%%%%%%%%%%%%%%%%%%%%%%%%%%%%%%%%%%%%%%%%%%%%%%%%%%%%%%%%%

%%%%%%%%%%%%%%%%%%%%%%%%%%%%%%%%%%%%%%%%%%%%%%%%%%%%%%%%%%%%%%%
%%%%%%%%%%%%%%%%%%%%%%%%%%%%%%%%%%%%%%%%%%%%%%%%%%%%%%%%%%%%%

\subsubsection{When $f(u)=\mathcal{A}\,sech^2(u/\varrho)$}

When $f(u)=\mathcal{A}\,sech^2(u/\varrho)$ we have only plotted $|\mathcal{I}_{\varepsilon}^{GW}(\Delta E)|$ as a function of the dimensionless detector transition energy, in Fig. \ref{fig:Ie-Inf-Gsn-sech2-vDE}, considering the eternal switching. The choice of eternal switching is motivated by the previous discussions on Fig. \ref{fig:Ie-Inf-Gsn-sech2-vDE}, and thus the quantity $|\mathcal{I}_{\varepsilon}^{GW}(\Delta E)|$ denotes the concurrence. Fig. \ref{fig:Ie-Inf-Gsn-sech2-vDE} reconfirms similar characteristics for the concurrence for $f(u)=\mathcal{A}\,sech^2(u/\varrho)$. Here also, the harvesting decreases with increasing transition energy and the distance between the static detectors. Larger $\varrho$ corresponds to higher harvesting in a low transition energy regime. While in a moderately high transition energy regime, lower $\varrho$ corresponds to higher harvesting. Thus both of our considered GW burst profiles (the Gaussian and sech-squared) exhibit similar characteristics in the entanglement harvesting measure. Furthermore, one can always harvest entanglement in this GW burst metric for the eternal switching scenario. Interestingly, in that case, the contribution from the flat background and due to the switching is zero. Then the harvesting solely happens due to the GW in the background spacetime. This situation is critical on its own for the burst type of GW, as we will see in Appendix \ref{Appn:IeGW-periodic-GWM}; similar things with periodic memory are somewhat uncertain.

\subsubsection{When $f(u)=\mathcal{A}\,\theta(u)$}

%%%%%%%%%%%%%%%%%%%%%%%%%%%%%%%%%%%%%%%%%%%%%%%%%%%%%%%%%%%%%
\begin{figure*}
\centering
\includegraphics[width=0.47\linewidth]{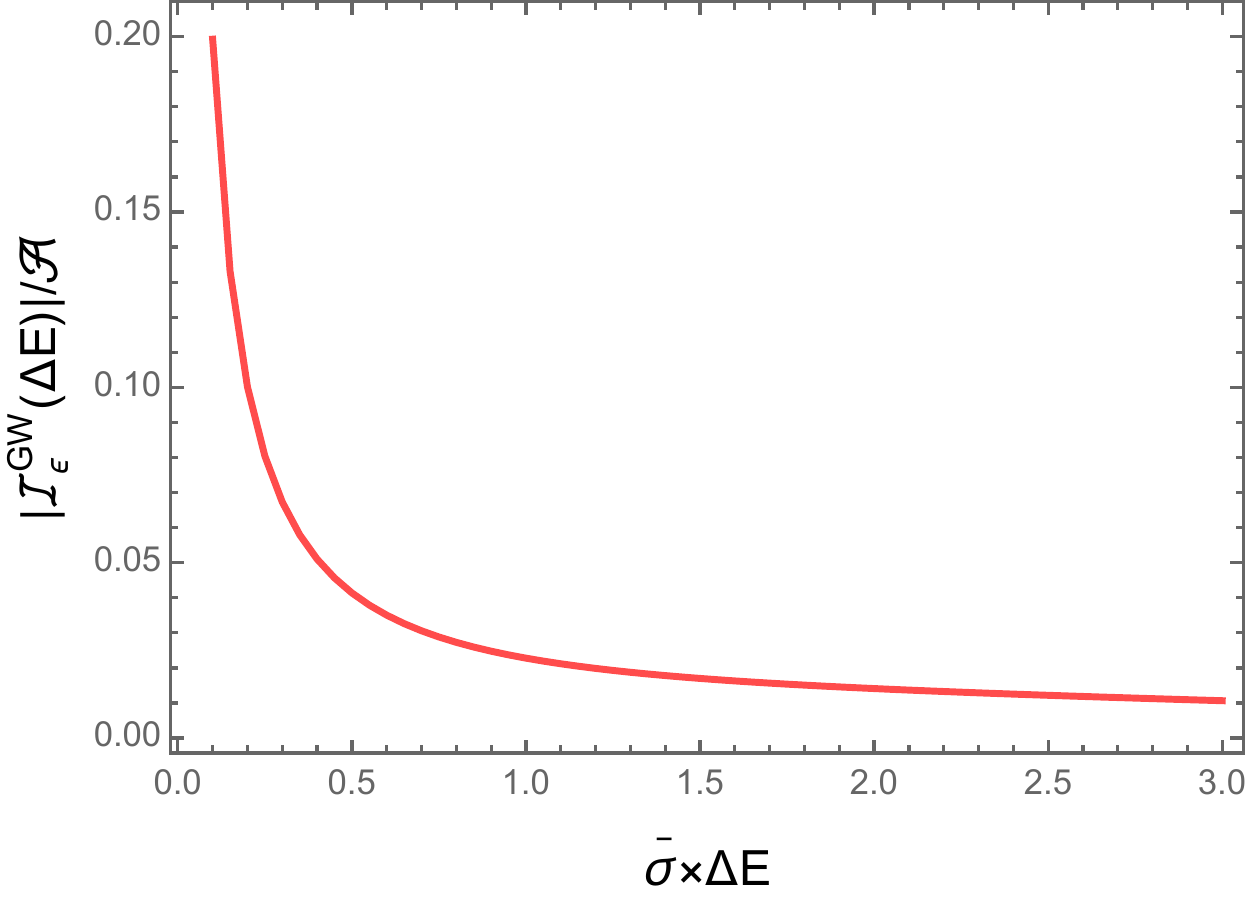}
\hskip 10pt
\includegraphics[width=0.47\linewidth]{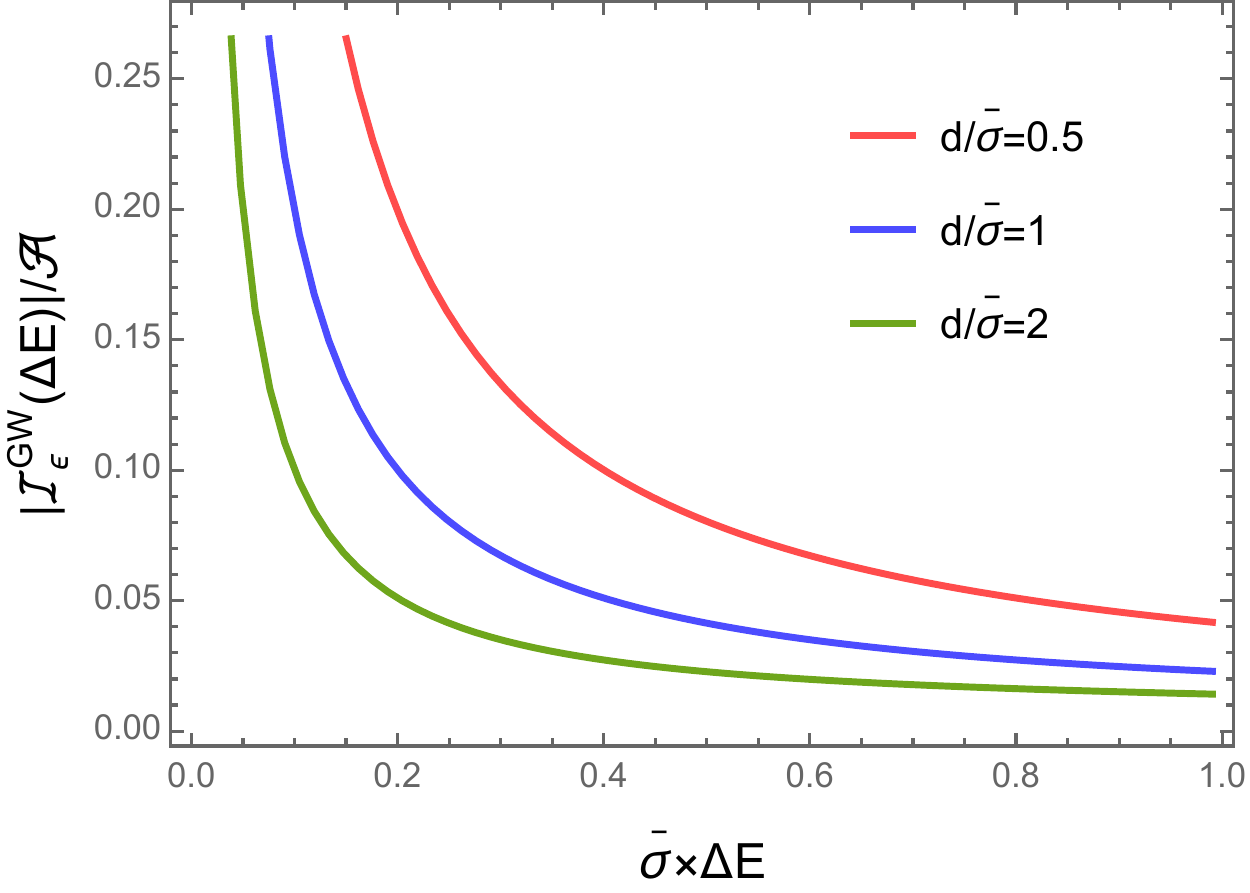}
\caption{  {The quantity $|\mathcal{I}_{\varepsilon}^{GW}(\Delta E)|$ is plotted as a function of the dimensionless detector transition energy $(\bar{\sigma}\,\Delta E)$ for $f(u)=\mathcal{A}\,\theta(u)$. We have considered the detectors interacting with the background field for infinite time, i.e., switching $\kappa(\tau_{j})=1$. Therefore, the quantity $|\mathcal{I}_{\varepsilon}^{GW}(\Delta E)|$ is the same as concurrence $\mathcal{C}_{\mathcal{I}}$. In the left plot, $d/\Bar{\sigma}=1$ is fixed. In the right plot, different curves correspond to different distances $d/\Bar{\sigma}$ between the static detectors. Here we observe that  entanglement harvesting is always larger for a lesser distance between the two detectors, and harvesting decreases with increasing transition energy. Unlike the Gaussian and sech-squared profiles of Fig. \ref{fig:Ie-Inf-Gsn-sech2-vDE}, here, as the transition energy decreases, the concurrence does not approach a fixed value but keeps increasing. Similar behavior is also observed with the tanh profile, see Fig. \ref{fig:Ie-Inf-tanh-vDE}.}}\label{fig:Ie-Inf-thetaFn-vDE}
\end{figure*}
%%%%%%%%%%%%%%%%%%%%%%%%%%%%%%%%%%%%%%%%%%%%%%%%%%%%%%%%%%%%%%

Finally, we consider a burst profile $f(u)=\mathcal{A}\,\theta(u)$, that has the general feature of bursts with memory. However, this burst is instantaneous and does not contain a time scale. In Fig. \ref{fig:Ie-Inf-thetaFn-vDE}, we have plotted the concurrence for eternal switching $\kappa(\tau_{j})=1$. To obtain these plots, we have utilized Eq. (\ref{eq:Ie-GW-thetaFn-Infinite-1}), and these plots have the salient features mentioned below:
%%%
\begin{itemize}
    \item From Fig. \ref{fig:Ie-Inf-thetaFn-vDE}, one can observe that the concurrence decreases with increasing detector transition energy and distance between the detectors.

    \item Here, the low transition energy behavior of concurrence differs from the burst profiles obtained in Fig. \ref{fig:Ie-Inf-Gsn-sech2-vDE}, i.e., without the memory contribution. However, this behavior is similar to the tanh profile curves that we will obtain in Fig. \ref{fig:Ie-Inf-tanh-vDE}. We observe that as the transition energy decreases, the concurrence keeps increasing, which is unlike the Gaussian and sech-squared profiles.
    
    \item As mentioned earlier for bursts with memory, we find that the integrand inside of expression of $\mathcal{I}_{ \varepsilon}^{GW}(\Delta E)$ from Eq. (\ref{eq:Ie-GW-thetaFn-Infinite-1}) behaves as $\mathcal{O}(1/\Delta E)$ as $\Delta E\to 0$, and this is the mathematical reason behind increasing concurrence with decreasing transition energy. This has to do with the universal feature of memory effect which in the Fourier space shows up as a pole in the zero frequency limit \cite{Strominger:2017}.

\end{itemize}

\color{black}

\subsubsection{When $f(u)=\mathcal{A}\,\{1+tanh(u/\lambda)\}$}

%%%%%%%%%%%%%%%%%%%%%%%%%%%%%%%%%%%%%%%%%%%%%%%%%%%%%%%%%%%%%
\begin{figure*}
\centering
\includegraphics[width=0.47\linewidth]{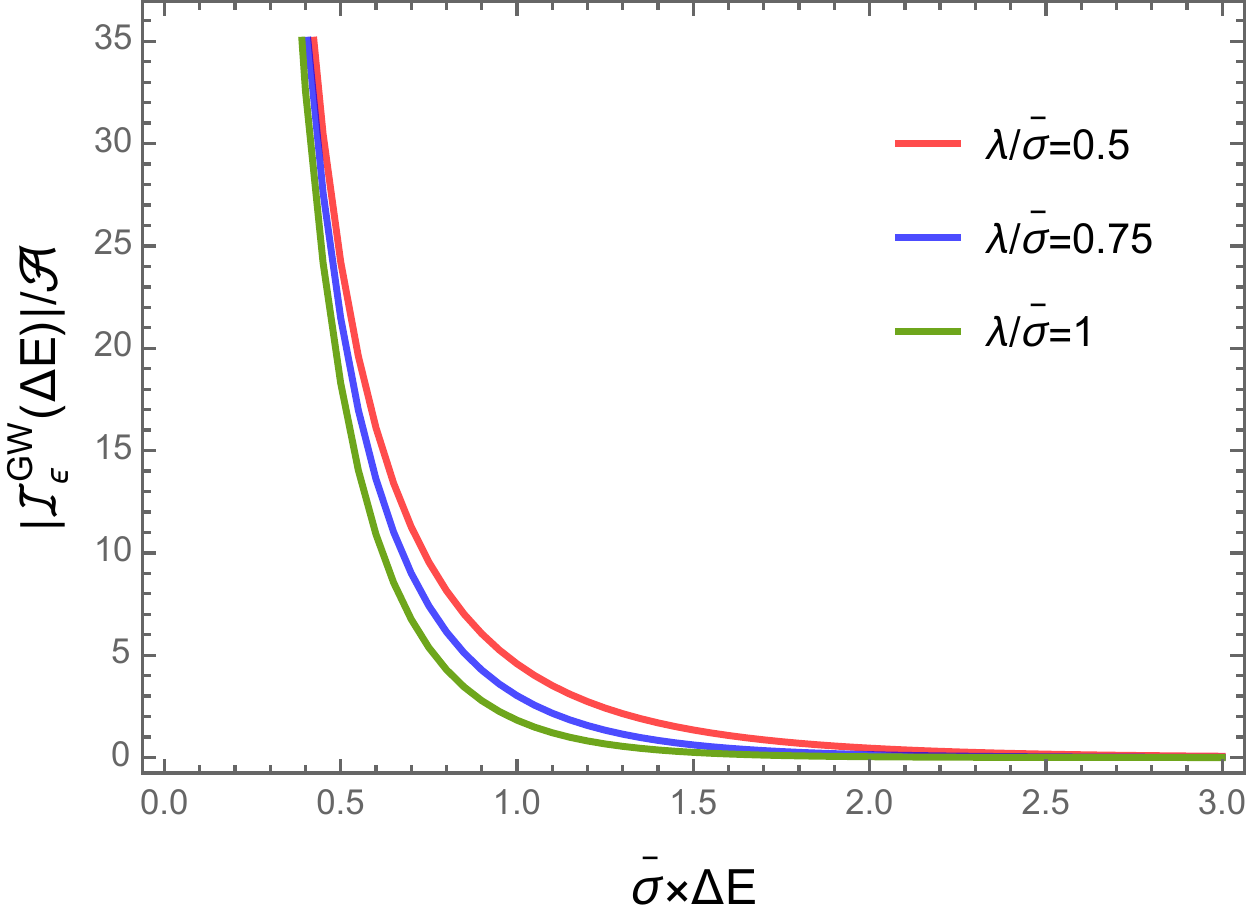}
\hskip 10pt
\includegraphics[width=0.47\linewidth]{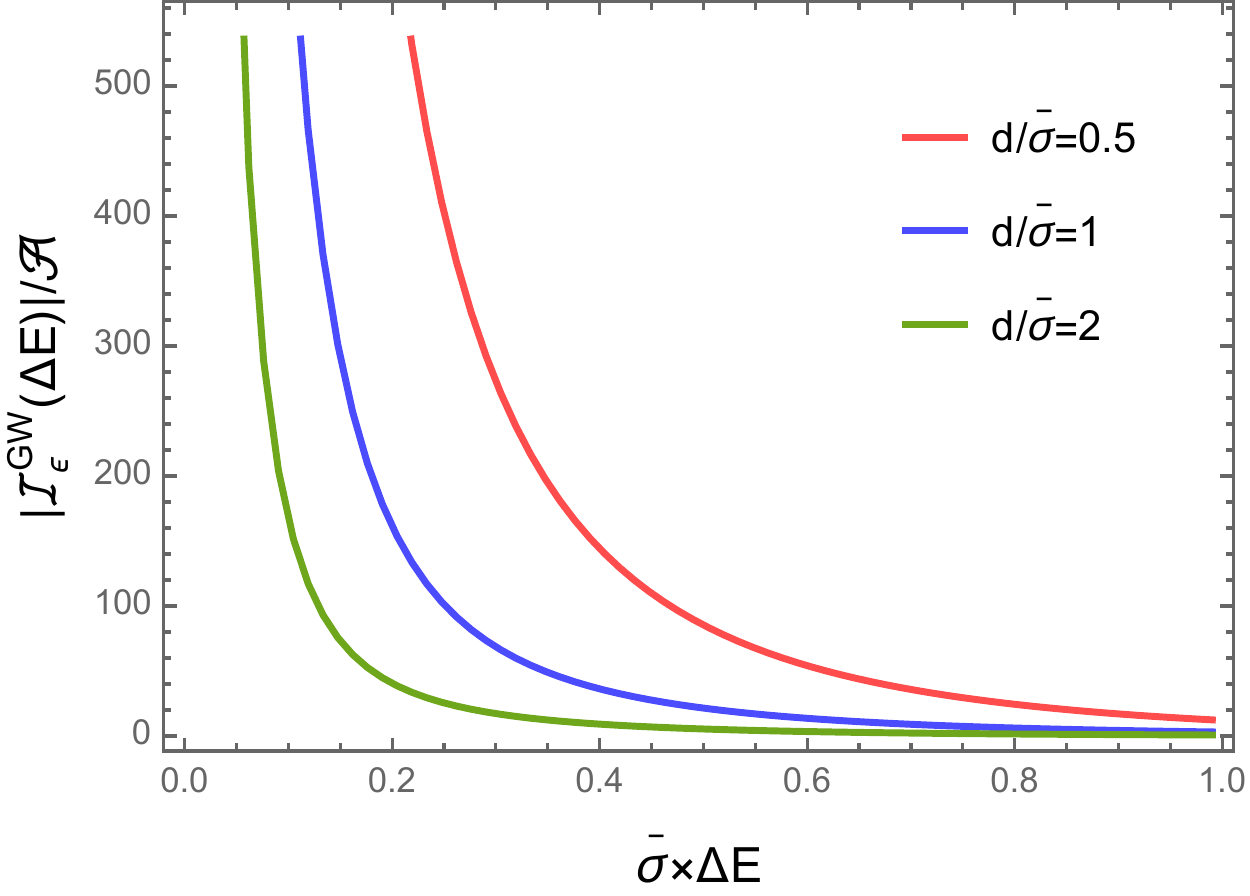}
\caption{The quantity $|\mathcal{I}_{\varepsilon}^{GW}(\Delta E)|$ is plotted as a function of the dimensionless detector transition energy $(\bar{\sigma}\,\Delta E)$ for $f(u)=\mathcal{A}\,\{1+tanh{(u/\lambda)}$. We have considered the detectors interacting with the background field for infinite time, i.e., with switching function $\kappa(\tau_{j})=1$. Therefore, the quantity $|\mathcal{I}_{\varepsilon}^{GW}(\Delta E)|$ is the same as concurrence. In the left plot, different curves correspond to different $\lambda/\Bar{\sigma}$, while $d/\Bar{\sigma}=1$ is fixed. In the right plot, different curves correspond to different distances $d/\Bar{\sigma}$ between the static detectors, and we have fixed $\lambda/ \Bar{\sigma}= 0.75$. Here we observe that one gets more harvesting for lower $\lambda/\Bar{\sigma}$. Furthermore, entanglement harvesting is always larger for a lesser distance between the two detectors, and harvesting decreases with increasing transition energy. The low energy qualitative natures of these curves are different than the ones obtained in Fig. \ref{fig:Ie-Gsn-vDE} and \ref{fig:Ie-Inf-Gsn-sech2-vDE}.}\label{fig:Ie-Inf-tanh-vDE}
\end{figure*}
%%%%%%%%%%%%%%%%%%%%%%%%%%%%%%%%%%%%%%%%%%%%%%%%%%%%%%%%%%%%%%

We now consider another burst profile with nonvanishing GW memory. In Fig. \ref{fig:Ie-Inf-tanh-vDE}, we have plotted the quantity $|\mathcal{I}_{\varepsilon}^{GW}(\Delta E)|$, which denotes the concurrence as well. To derive this expression, we employ Eq. (\ref{eq:Ie-GW-tanh-Infinite-4}) and assume the switching function to be $\kappa(\tau_{j})=1$. The salient features are summarised below:
%%%
\begin{itemize}
    \item From Fig. \ref{fig:Ie-Inf-tanh-vDE}, one can observe that the concurrence decreases with increasing detector transition energy and distance between the detectors.

    \item Here, the low transition energy behavior of concurrence differs from the burst profiles obtained in Fig. \ref{fig:Ie-Gsn-vDE} and \ref{fig:Ie-Inf-Gsn-sech2-vDE}, i.e., without the memory contribution.  However, this behavior is similar to the concurrence obtained from Fig. \ref{fig:Ie-Inf-thetaFn-vDE} for step function profile.

   \item We also confirm from Fig. \ref{fig:Ie-Inf-tanh-vDE} that at a very low detector transition energy regime of $\Delta E\to 0$, the concurrence does not tend to reach a finite value, unlike the Gaussian and sech-squared GW profiles. 

    \item One should also note from Fig. \ref{fig:Ie-Inf-tanh-vDE} that the concurrence decreases with increasing $\lambda/\Bar{\sigma}$.
    
\end{itemize}
%%%

\section{Non-geodesic detectors in Minkowski background}\label{sec:Detectors-trajectories-Minkowski}

In this section, we consider specific non-geodesic trajectories for the detectors in the Minkowski background. More specifically, we want to discover what happens if these trajectories mimic the geodesic trajectories in the GW burst backgrounds. In this regard, we shall first find out the geodesic trajectories in the background of the metric (\ref{eq:BJR-metric}). With a proper time $\tau$, the geodesic equations for the velocity components $U^{\mu}$ will be 
\begin{eqnarray}\label{eq:Dtj-Minkowski-1}
    dU^{u}/d\tau &=& 0~,\nonumber\\
    dU^{x}/d\tau &=& -U^{u}U^{x}f'(u)~,\nonumber\\
    dU^{y}/d\tau &=& U^{u}U^{y}f'(u)~,\nonumber\\
    dU^{v}/d\tau &=& (U^{y^2}-U^{x^2})f'(u)~;
\end{eqnarray}
where $f'(u)$ denotes the derivative of $f(u)$ with respect to $u$. From these equations, one can easily find out the velocity components as:
\begin{eqnarray}\label{eq:Dtj-Minkowski-2}
    U^{u} &=& C_{u}~,\nonumber\\
    U^{x} &=& C_{x}\,e^{-f(u)}~,\nonumber\\
    U^{y} &=& C_{y}\,e^{f(u)}~,\nonumber\\
    U^{v} &=& C_{v}+\frac{1}{2\,C_{u}}\,\Big[C_{y}^2\,e^{2f(u)}+C_{x}^2\,e^{-2f(u)}\Big]~;
\end{eqnarray}
where $C_{u}$, $C_{x}$, $C_{y}$, and $C_{v}$ are integration constants. One can integrate again the velocities $U^{\mu} = dx^{\mu}/d\tau$ to find out the coordinates $x^{\mu}$ in the geodesic trajectories. We find these trajectories to be 
\begin{eqnarray}\label{eq:Dtj-Minkowski-3}
    u &=& C_{u}\,\tau+\tilde{C}_{u}~,\nonumber\\
    x &=& \frac{C_{x}}{C_{u}}\,\int e^{-f(u)}\,du + \tilde{C}_{x}~\approx \frac{C_{x}}{C_{u}}\, \big[u-\mathcal{A}\, \bar{g}(u)\big] + \tilde{C}_{x}~,\nonumber\\
    y &=& \frac{C_{y}}{C_{u}}\,\int e^{f(u)}\, du + \tilde{C}_{y}~ \approx \frac{C_{y}}{C_{u}}\, \big[u+\mathcal{A}\, \bar{g}(u)\big] + \tilde{C}_{y}~,\nonumber\\
    v &=& \frac{C_{v}}{C_{u}}\,u+\frac{1}{2\,C_{u}^2}\,\Big[C_{y}^2\,\int e^{2f(u)} du +C_{x}^2\,\int e^{-2f(u)} du\Big]+\tilde{C}_{v}~\nonumber\\
    &\approx& \frac{C_{v}}{C_{u}}\,u+\frac{1}{2\,C_{u}^2}\,\Big[C_{y}^2\,\big[u+2\mathcal{A}\, \bar{g}(u)\big] +C_{x}^2\,\big[u-2\mathcal{A}\, \bar{g}(u)\big]\Big]+\tilde{C}_{v}~.
\end{eqnarray}
Here $\{\tilde{C}_{\mu}\}$ is another set of constants of integration. We have considered $f(u)=\mathcal{A}\,g(u)$, $\Bar{g}(u)=\int g(u)\,du$, and a very small perturbation strength $\mathcal{A}$, i.e., $\mathcal{A}\ll 1$, to get the second expressions in Eq. (\ref{eq:Dtj-Minkowski-3}).

\iffalse
%
\begin{eqnarray}\label{eq:Dtj-Minkowski-4}
    u &=& C_{u}\,\tau+\tilde{C}_{u}~,\nonumber\\
    x &\approx& \frac{C_{x}}{C_{u}}\, \big[u-\mathcal{A}\, \bar{g}(u)\big] + \tilde{C}_{x}~,\nonumber\\
    y &\approx& \frac{C_{y}}{C_{u}}\, \big[u+\mathcal{A}\, \bar{g}(u)\big] + \tilde{C}_{y}~,\nonumber\\
    v &\approx& \frac{C_{v}}{C_{u}}\,u+\frac{1}{2\,C_{u}^2}\,\Big[C_{y}^2\,\big[u+2\mathcal{A}\, \bar{g}(u)\big] +C_{x}^2\,\big[u-2\mathcal{A}\, \bar{g}(u)\big]\Big]+\tilde{C}_{v}~.
\end{eqnarray}
%
\fi

Now let us consider specific values for the integration constants so that we can get the relevant trajectories to perceive some expression for the Minkowski-Wightman function similar in form to the expression from Eq. (\ref{eq:Wightman-fn-Gaussian-2}). For detector $A$, we consider the trajectory such that $C_{u}=C_{x}=C_{y}=1$, and $C_{v}=\tilde{C}_{u}=\tilde{C}_{x}=\tilde{C}_{y}=\tilde{C}_{v}=0$. Whereas, for detector $B$, we consider the trajectory in a way such that $C_{u}=C_{x}=C_{y}=1$, $C_{v}=\tilde{C}_{u}= \tilde{C}_{y}= \tilde{C}_{v}=0$, and $\tilde{C}_{x}=d$. Then from Eq. (\ref{eq:Dtj-Minkowski-3}) we get the trajectory of detector $A$ as: $u_{A}=\tau_{A}$, $x_{A} = u_{A}-\mathcal{A}\, \bar{g}(u_{A})$, $y_{A} = u_{A}+\mathcal{A}\, \bar{g}(u_{A})$, and $v_{A} = u_{A}$. On the other hand, we get the trajectory of detector $B$ as $u_{B}=\tau_{B}$, $x_{B} = u_{B}-\mathcal{A}\, \bar{g}(u_{B})+d$, $y_{B} = u_{B}+\mathcal{A}\, \bar{g}(u_{B})$, and $v_{B} = u_{B}$. Naturally, the quantity $(\sigma_{M}+i\,\epsilon\,\Delta u)$ in the denominator of the Minkowski Wightman function from Eq. (\ref{eq:Wightman-fn-Gaussian-2-M}) will now become $(\sigma_{M}+i\,\epsilon\,\Delta u)\approx (\Delta\tau-d)^2+i\,\epsilon\,\Delta\tau+2\,d\,\mathcal{A}\,\big[\bar{g}(\tau_{A})-\bar{g}(\tau_{B})\big]$, with $\Delta\tau=\tau_{A}-\tau_{B}$. The whole Minkowski-Wightman function can be approximated for $\mathcal{A}\ll 1$ to be
\begin{eqnarray}\label{eq:Dtj-Minkowski-5}
    G_{W_{M}}(\tau_{A},\tau_{B}) &\approx& \frac{1}{4\pi^2}\times\frac{1}{(\Delta\tau-d)^2+i\,\epsilon\,\Delta\tau}~\nonumber\\
    ~&&~ + ~ \frac{1}{2\pi^2}\times\frac{\mathcal{A}\,\big\{\bar{g}(\tau_{A})-\bar{g}(\tau_{B})\big\}}{\Big[(\Delta\tau-d)^2+i\,\epsilon\,\Delta\tau\Big]^2}~.
\end{eqnarray}
Here the second term looks similar to $G_{W_{GW}}$ from Eq. (\ref{eq:Wightman-fn-Gaussian-2-GW}), which arrives from the gravitational wave contribution. However, this similarity is only in the numerator, while the denominator has a different structure. In this case, the poles are obtained from $(\Delta\tau-d)^2+i\,\epsilon\,\Delta\tau=0$, whereas the poles in our previous case of Eq. (\ref{eq:Wightman-fn-Gaussian-2}) were obtained from $\Delta\tau^2-d^2-i\,\epsilon\,\Delta\tau=0$. Therefore, it is evident that the results from this current consideration will not be the same as our main results.

We should mention again that in our main estimations, we considered the detectors to be static in the GW background. Therefore the motions of the detectors have little contribution to entanglement harvesting, and one perceives that the harvesting profiles from Sec. \ref{sec:entanglement-harvesting} are entirely influenced by the background. Notably, in \cite{VerSteeg:2007}, it is observed that entanglement measures can distinguish between a thermal bath and de Sitter spacetime, which single detector transition is unable to provide. It would have also been interesting if our initial consideration for the detectors were in geodesic trajectories. In that scenario also, we could have compared those results with the ones obtained in this section and investigated how the motion of detectors and curvature distinctly affect entanglement harvesting.

\color{black}

\section{Discussion}\label{sec:discussion}

In this work, we have investigated the entanglement  harvesting condition and the measure of harvested entanglement between two static Unruh-DeWitt detectors in backgrounds of GW bursts with and without memory. {  The burst profiles without memory are studied to compare the generic features of the burst with memory {\em vis-a-vis} entanglement harvesting. For GW burst without memory, we considered the Gaussian and sech-squared type profiles; and for GW burst with memory, we chose the tanh and Heaviside step function type profiles.} We find that entanglement harvesting is possible for all these burst profiles, including and excluding GW memory. However, memory seems to introduce qualitative differences in entanglement measurements, for example, in the value of concurrence.

Let us go through the main observations and elucidate these issues.

\begin{itemize}
    \item We observed characteristic similarities between the concurrences from the two burst profiles without memory. This is the primary reason for working with two separate cases of burst profiles without memory. One can clearly find that the measures of entanglement are alike in such symmetric pulses. For these profiles, we observed lengthy bursts provide greater entanglement harvesting at low detector transition energies (see Fig. \ref{fig:Ie-Inf-Gsn-sech2-vDE}). While for relatively higher detector transition energies, shorter bursts give higher harvesting (Figs. \ref{fig:Ce-Gsn-vDE} and \ref{fig:Ie-Inf-Gsn-sech2-vDE}).

    \item On the other hand, for a GW burst with memory from the tanh-type profile, it is inferred that shorter bursts always provide greater harvesting \ref{fig:Ie-Inf-tanh-vDE}. With the Heaviside step function profile, the burst is instantaneous and there is no way to understand the harvesting dependence for varying burst duration. However, in this scenario of Fig. \ref{fig:Ie-Inf-thetaFn-vDE} also we observe that the harvesting profile is similar to the tanh one of Fig. \ref{fig:Ie-Inf-tanh-vDE}. In both the Heaviside step function and tanh cases, the harvesting keeps increasing as the transition energy decreases, unlike the symmetric pulse cases where the harvesting tends to reach finite values.

    \item Our observations suggest GW bursts without memory act somewhat like the switching functions. When the burst is short, it provides additional excitation to the quantum state of the field. Therefore, one reports greater harvesting with shorter bursts in large detector transition energy regimes. Noticeably, this particular observation in a higher transition energy regime is true for all the considered burst profiles with or without memory. However, when the detector transition energy is small, the differences between the bursts with and without memory are prominent in the harvesting profile. Naturally, the small transition energy regime is important to distinguish between the two types of burst profiles. Moreover, if one considers a static detector in a Minkowski vacuum with the Gaussian switching, the qualitative behavior of the non-local entangling term $|\mathcal{I}_{\varepsilon}^{M}|$ (expression of Eq. (\ref{eq:Ie-M-Gaussian-2}), see Fig. \ref{fig:Ie-Gaussian-Mink-static} of Appendix \ref{Appn:Ie-term-Minkowski}) would be similar to the one observed in Fig. \ref{fig:Ie-Inf-Gsn-sech2-vDE} for the Gaussian and sech-squared profiles. In Fig. \ref{fig:Ie-Gaussian-Mink-static} we observe in a low transition energy regime, larger interaction time corresponds to larger $|\mathcal{I}_{\varepsilon}^{M}|$, while for higher transition energy, shorter interaction time corresponds to higher $|\mathcal{I}_{\varepsilon}^{M}|$, compare Figs. \ref{fig:Ie-Inf-Gsn-sech2-vDE} and \ref{fig:Ie-Gaussian-Mink-static}.

    \item However, we have checked, and a similar analogy cannot be made for the GW bursts with memory (Heaviside step function and tanh profiles) with similar switching. For example, if one considers a Heaviside step function switching, i.e., for $\kappa(\tau_{j})=\theta(\tau_{j})$, the characteristics of the Minkowski non-local term $\mathcal{I}_{\varepsilon}^{M}$ would not be similar to the plots from Figs. \ref{fig:Ie-Inf-thetaFn-vDE} and \ref{fig:Ie-Inf-tanh-vDE}, see Eq. (\ref{eq:Ie-Mink-static-ThetaFn}) from Appendix \ref{Appn:Ie-term-Minkowski} and the discussion therein. Let us elucidate the characteristics of these two figures by drawing a comparison with the scenario without memory. For the case without memory (Fig. \ref{fig:Ie-Inf-Gsn-sech2-vDE}) and in a very small transition energy regime, as one  considers the smaller and smaller duration of the burst, the amplitude of entanglement harvesting will keep decreasing. Moreover, for an infinitesimal duration of the burst, the amplitude of the entanglement harvesting will become minuscule, i.e., in this scenario, the harvesting profile does not report any change in spacetime and sees it to be effectively flat. Whereas, in the low transition energy regime, even if one considers a very small duration of memory burst, even for instantaneous bursts like the Heavisede step function, there will be entanglement harvesting, see Figs. \ref{fig:Ie-Inf-tanh-vDE} and \ref{fig:Ie-Inf-thetaFn-vDE}. This signifies that the harvesting profile is sensitive to the memory bursts, and in this scenario, it senses that the spacetime property is changed, and the situation cannot be compared naively with a static detector from the Minkowski background with non-trivial switching like without memory bursts.

    \item A typical feature of the bursts with memory is that it falls of as either $\mathcal{O}(1/\Delta E)$ or $\mathcal{O}(1/\Delta E^2)$ in the limit $\Delta E\to 0$. The step-function profile (which is taken as the standard memory profile in the GW physics literature) reproduces a term similar to the Weinberg leading soft factor \cite{Strominger:2017}. As argued previously, this occurs because the formula of the integral $\mathcal{I}_{\varepsilon}^{GW}$ is almost like a Fourier transform of the gravitational pulse profile. We expect, in general, the concurrence measure to grow rapidly as $\Delta E\to 0$ since the memory profiles would qualitatively mimic the mathematical nature of the step function/tanh-function. The tanh-function is analyzed in our paper as it is a smooth function and one gets a burst profile with varying duration.

    \item Furthermore, for all of the considered burst profiles, we observed entanglement harvesting increases with decreasing distance between the two detectors and with decreasing detector transition energy. 

    \item Our entire work considers static Unruh-deWitt detectors. Thus, detector motion does not affect the difference in the harvested entanglement for burst profiles with and without memory. Whatever difference appears in the measure of the harvested entanglement comes from the spacetime geometry corresponding to the types of burst profiles. This is because, in bursts with and without memory, the metric function containing pulse profiles ($f(u)$ in Eq. (\ref{eq:BJR-metric})) are mathematically different. Also, note that the scalar field behavior at asymptotic past and future also changes for bursts with memory.

    \item We have also checked in Sec. \ref{sec:Detectors-trajectories-Minkowski} how non-geodesic detector trajectories in Minkowski spacetime mimicking the geodesics in plane gravitational wave metric provide different Wightman function and should result in different concurrence measures compared to the current results. Specifically, we find that the pole structures of the Wightman functions are different in the two cases. Hence, it provides additional support to the claim that the entanglement harvesting in the present scenario depends on spacetime geometry.

\end{itemize}
\vspace{0.1cm}
\color{black}

Let us also briefly mention the major distinctions of our findings from the previously studied $f(u)=\mathcal{A}\,\cos{(\omega u)}$ type GW profiles \cite{Xu:2020pbj}. In the latter case, entanglement harvesting has a resonance-like effect in it whenever the detector transition energy matches the wave frequency, a behavior not obtainable from our considered burst profiles. Moreover, the behavior of the specific non-local entangling term $|\mathcal{I}_{\varepsilon}^{GW}(\Delta E)|$ from these $\cos{(\omega u)}$ type profiles with the Gaussian switching has some distinct qualitative features compared to the non-local terms from our considered GW burst profiles. In this regard, see Fig. \ref{fig:Ie-Gsn-cos-vDE} and compare it with the Figs. \ref{fig:Ie-Gsn-vDE} and \ref{fig:Ie-Inf-Gsn-sech2-vDE}. In particular, we observe that in the low detector transition energy regimes, the modulus of that non-local term from $\cos{(\omega u)}$ type profile is larger for a smaller frequency of the GW. At the same time, that non-local term's modulus is larger for higher frequency $\omega$ for larger detector transition energies. However, we should mention that here also $|\mathcal{I}_{\varepsilon} ^{GW} (\Delta E)|$ increases with decreasing distance between the two detectors. Moreover, with eternal switching, one can always harvest entanglement from our considered GW backgrounds, i.e., for the Gaussian, sech-squared, and tanh profiles, with arbitrary detector transition energies. However, the same cannot be claimed true for the periodic profile; see Appendix \ref{Appn:IeGW-cos-Infty} and the discussions therein. These observations specify the characteristic differences in the entanglement harvesting profiles between the periodic and our considered GW burst profiles.\vspace{0.1cm}

Finally, we point out that understanding the qualitative and quantitative dependence of entanglement harvesting on GW memory is still due. The setup considered in this work deals with two static detectors ({\em i.e.} spacetime trajectories are kept fixed via non-gravitational interaction).  Even after the passage of a GW, the positions remain fixed. In consequence, the deviations in their trajectories, i.e., the imprint of the GW memory, is not captured in the estimations of entanglement harvesting. Specifically, the harvesting characteristics observed due to these GW bursts have their origin in the deviation of these backgrounds from the Minkowski spacetime. It is an important observation, as static detectors would not have harvested any entanglement in the Minkowski background \cite{Koga:2018the}. Furthermore, it will be interesting to construct setups where the detectors can sense the passage of the GWs \cite{Chen:2021}, i.e., their trajectories get altered due to the passing wave. Then there is a possibility that the GW memory encoded in the detectors' trajectories can also get reflected in the harvested entanglement. That will provide a procedure for interpreting the background memory from the harvested entanglement. We are currently exploring this particular direction and hope to address them in future communication.

\iffalse
Finally, we shall like to mention that the setup considered here deals with two static detectors, where the spacetime is already modified due to the passage of a GW burst. However, a more realistic scenario would be to consider a dynamical background.

  {Write about the possible future directions, like how the passing of GW may affect the entanglement dynamics.} \fi

\begin{acknowledgments}

We thank Sumanta Chakraborty and Sk Jahanur Hoque for useful discussions. We also thank Bibhas Ranjan Majhi for his comments on the manuscript. The authors would also like to acknowledge the anonymous referee whose insightful comments helped in enhancing the quality of the paper. S.B. would like to thank the Science and Engineering Research Board (SERB), Government of India (GoI), for supporting this work through the National Post Doctoral Fellowship (N-PDF, File number: PDF/2022/000428). I.C. would like to thank Amit Ghosh for arranging a visit in the Theory Division of Saha Institute of Nuclear Physics, Kolkata, India, where a significant portion of the work was carried out. I.C. also thanks Indian Institute of Technology Bombay (IIT Bombay) for supporting this work through a postdoctoral fellowship.  S.M. acknowledges the financial support from the fellowship Lumina Quaeruntur No. LQ100032102 of the Czech Academy of Sciences.

\end{acknowledgments}

\appendix

\begin{widetext}

%%%%%%%%%%%%%%%%%%%%%%%%%%%%%%%%%%%%%%%%%%%%%%%%%%%%%%%%%%%%%%%%%%%%%%%%%%%%%
 
\section{Discussion on the consistency of the perturbative approach}\label{Appn:perturbation-consistency}

In this part of the appendix, we discuss the validity of the perturbative approach adopted in this work and in this particular scenario of GW burst backgrounds. It would have been best if one could obtain the mode functions and the Wightman functions exactly without using the perturbative approach and then numerically evaluate the concurrence by integrating these Wightman functions. Then compare these results with the perturbative approach to check its validity. However, it is not possible to do that with different bust profiles. Therefore, the next plausible thing to do for checking the validity of this perturbative approach is to investigate the higher order terms, i.e., $\mathcal{O}(\mathcal{A}^2)$, in the mode functions and the Witghman functions. Let us understand a few things in this direction.

The mode functions for a general linear gravitational wave perturbation $f(u)=\mathcal{A}\,g(u)$, with $\mathcal{A}$ being the perturbation strength, is given by
\begin{eqnarray}\label{eq:modeFn-general}
u_{\mathbf{k}}(X) &\simeq& \frac{1}{\sqrt{2k_{-}(2\pi)^3}}\,e^{-i\,k_{-}v+i\,k_{1}x+i\,k_{2}y-i\frac{(k_{1}^2+k_{2}^2)}{4k_{-}}u} \nonumber\\
~&\times& \exp{\bigg[\frac{i\,\mathcal{A}}{4k_{-}}(k_{1}^2-k_{2}^2)\,\Bar{g}(u)\bigg]}\,,
\end{eqnarray}
where $\Bar{g}(u)=\int g(u)\,du$, see Eqs. (\ref{eq:mode-fn-Gaussian}), (\ref{eq:mode-fn-sech2}), (\ref{eq:mode-fn-tanh}), and (\ref{eq:mode-fn-thetaFn}) and \cite{Garriga-1991}. For the perturbation strength to be very small, i.e., for $\mathcal{A}\ll 1$, we previously considered terms up to $\mathcal{O}(A)$ in the evaluation of $G_{W_{GW}}(X,X')$. Let us now consider terms $\mathcal{O}(A^2)$ to evaluate the same and check whether the perturbation in $\mathcal{A}$ remains consistent to evaluate the entanglement measures. Let us consider $G_{W_{GW}}^{2}(X,X')$ to be the part of the Wightman function that contains $\mathcal{O}(A^2)$ term. This part of the Wightman function is evaluated from
\begin{eqnarray}\label{eq:GreensFn-A-square-1}
G_{W_{GW}}^{2}(X,X') &=& -\frac{\mathcal{A}^2}{32}\int_{0}^{\infty}\frac{dk_{-}}{2k_{-}(2\pi)^3}\int_{-\infty}^{\infty}dk_{1}\,\int_{-\infty}^{\infty}dk_{2}\,e^{-i\,k_{-}\Delta v+i\,k_{1}\Delta x+i\,k_{2}\Delta y-i\frac{(k_{1}^2+k_{2}^2)}{4k_{-}}\Delta u} \nonumber\\
~&\times& \frac{(k_{1}^2-k_{2}^2)^2}{k_{-}^2}\Delta\Bar{g}^2(u,u')\,,
\end{eqnarray}
where $\Delta\Bar{g}(u,u')=\Bar{g}(u)-\Bar{g}(u')$. After carrying out the $k_{1}$ and $k_{2}$ integrations we shall get
\begin{eqnarray}\label{eq:GreensFn-A-square-2}
G_{W_{GW}}^{2}(X,X') &=&  \frac{i\,\mathcal{A}^2}{8 \pi ^2 \text{$\Delta $u}^5}\,
\Delta\Bar{g}^2(u,u')\,\int_{0}^{\infty}dk_{-}\,e^{-\frac{i k_{-} \left(\Delta u \,\Delta v-\Delta x^2-\Delta y^2\right)}{\Delta u}}\nonumber\\
~&\times& \bigg[\Delta u^2+2 i \Delta u\, k_{-} \left(\Delta x^2+\Delta y^2\right)-k_{-}^2 \left(\Delta x^2-\Delta y^2\right)^2\bigg]~,
\end{eqnarray}
One can evaluate this integral by introducing a regulator of the form $e^{-k_{-}\epsilon}$. The evaluation of the first two terms in the bracket is given previously; see Eq. (\ref{eq:Wightman-fn-Gaussian-2}). The integration with $k_{-}^2$ term in the bracket will be $\int_{0}^{\infty}dk_{-}\,k_{-}^2\,e^{i\,k_{-}(\sigma_{M}/\Delta u)}\, e^{-k_{-}\epsilon} =  2/(\sigma_{M}/\Delta u+i\,\epsilon)^3$. 

We should note that compared to (\ref{eq:Wightman-fn-Gaussian-2}) in the expression of (\ref{eq:GreensFn-A-square-2}), there are no suspicious terms coming out of the wave-number ($k_{1}$, $k_{2}$, and $k_{-}$) integration that may give rise to some diverging quantities which may render the perturbative approach invalid. We further notice that in the current scenario with $\mathcal{A}^2$ term we have $\Delta\Bar{g}^2(u,u')$ multiplied, and $\Delta\Bar{g}(u,u')$ was multiplied in (\ref{eq:Wightman-fn-Gaussian-2}). Then by our understanding, the form of this quantity $\Delta\Bar{g}^2(u,u')$, which depends on $f(u)$, can only dictate the incoming of any divergences in the concurrence that can make the perturbative approach invalid. In fact, for eternal switching, we will observe that periodic GW perturbations may result in divergences in the non-local entangling term, see Appendix \ref{Appn:IeGW-cos-Infty}. However, our considered GW perturbations are due to bursts, and they do not carry the characteristics of the periodic profile. Regardless to say as $\mathcal{A}\sim 10^{-21}$, for finite values of the integral in (\ref{eq:GreensFn-A-square-2}), the absolute value of $G_{W_{GW}}^{2}(X,X')$ should be much lesser than $G_{W_{GW}}(X,X')$. Therefore, even with eternal switching, our perturbative procedure will remain valid with the considered burst profiles.

\color{black}

\section{Non-local term in Minkowski vacuum with non-trivial switching}\label{Appn:Ie-term-Minkowski}

%%%%%%%%%%%%%%%%%%%%%%%%%%%%%%%%%%%%%%%%%%%%%%%%%%%%%%%
\begin{figure*}
\centering
\includegraphics[width=0.47\linewidth]{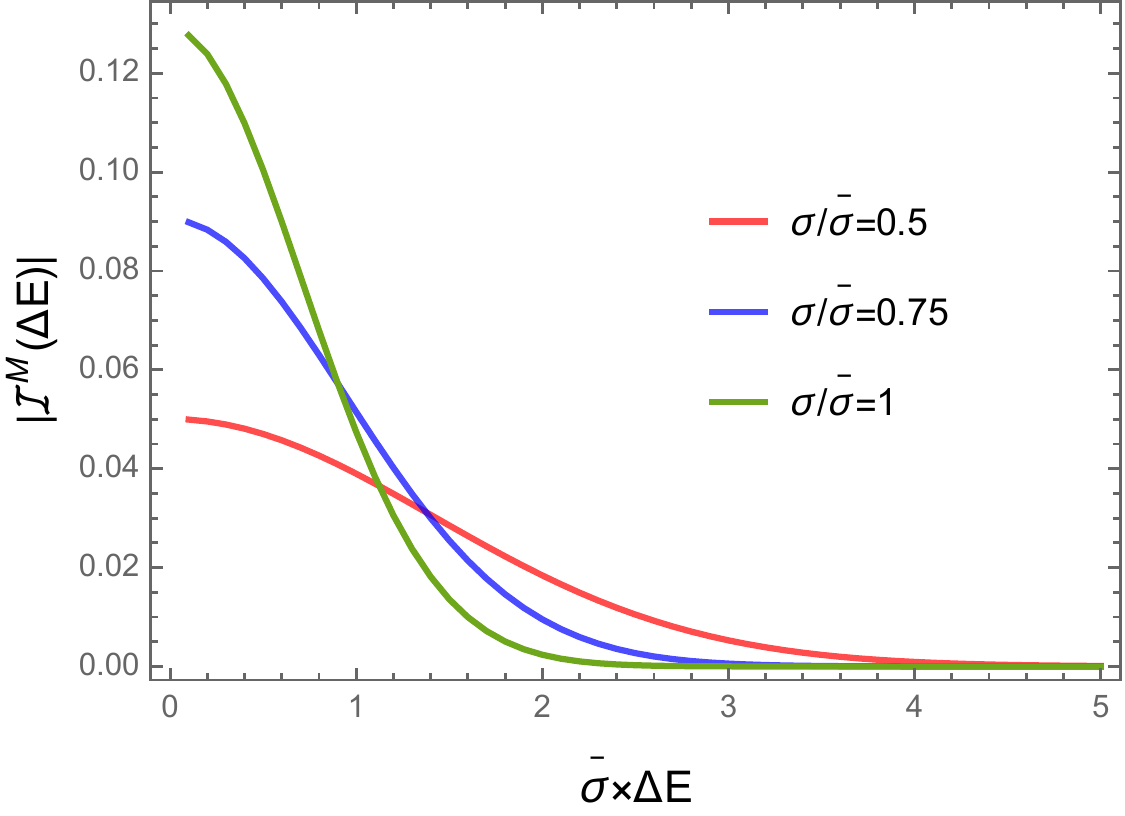}
\caption{  {The modulus of the quantity $\mathcal{I}_{\varepsilon}^{M}(\Delta E)$ is plotted as a function of the dimensionless detector transition energy $(\sigma\,\Delta E)$ for $f(u)=\mathcal{A} \cos{(\omega\,u)}$. The above plots correspond to detectors interacting with the background field with Gaussian switching functions $\kappa(\tau_{j})=e^{-\tau_{j}^2/(2\sigma^2)}$. Different curves correspond to different switching duration $\sigma$.}}\label{fig:Ie-Gaussian-Mink-static}
\end{figure*}
%%%%%%%%%%%%%%%%%%%%%%%%%%%%%%%%%%%%%%%%%%%%%%%%%%%%%%%%

In this section, we discuss the behavior of the modulus of the non-local term in the Minkowski vacuum for a static detector, i.e., $|\mathcal{I}_{\varepsilon}^{M}(\Delta E)|$, for the Gaussian switching. In this regard, we consider the expression of $\mathcal{I}_{\varepsilon}^{M}(\Delta E)$ from Eq. (\ref{eq:Ie-M-Gaussian-2}) and the modulus of this quantity is plotted   in Fig. \ref{fig:Ie-Gaussian-Mink-static}.\vspace{0.2cm}

Let us now evaluate $\mathcal{I}_{\varepsilon}^{M}$ from Eq. (\ref{eq:Ie-general-decomp}), but with a Heaviside step function switching $\kappa(\tau_{j})=\theta(\tau_{j})$. For two static detectors separated by a distance $d$, this Minkowski non-local term will be
\begin{widetext}
\begin{eqnarray}\label{eq:Ie-Mink-static-ThetaFn}
    \mathcal{I}_{\varepsilon}^{M} &=& \frac{1}{4\pi^2}\int_{-\infty}^{\infty}d\Bar{\eta}\int_{-\infty}^{\infty}d\Bar{\xi}\,e^{i\Delta E\,\Bar{\xi}}\frac{\theta(\Bar{\eta})}{\Bar{\eta}^2-d^2-i\,\Bar{\eta}\,\epsilon}\,\theta\bigg(\frac{\bar{\xi}+\bar{\eta}}{2}\bigg)\,\theta\bigg(\frac{\bar{\xi}-\bar{\eta}}{2}\bigg)~,\nonumber\\
    ~&=& \frac{1}{4\pi^2}\int_{0}^{\infty}d\Bar{\eta}\int_{-\Bar{\eta}}^{\Bar{\eta}}d\Bar{\xi}\,\frac{e^{i\Delta E\,\Bar{\xi}}}{\Bar{\eta}^2-d^2-i\,\Bar{\eta}\,\epsilon}~,\nonumber\\
    ~&=& \frac{1}{2\pi^2\,\Delta E}\int_{0}^{\infty}d\Bar{\eta}\,\frac{\sin{(\Delta E\,\Bar{\eta})}}{\Bar{\eta}^2-d^2-i\,\Bar{\eta}\,\epsilon}~.
\end{eqnarray}
\end{widetext}
In the limit of $\Delta E\to 0$, the integrand inside this integral becomes finite, unlike the expressions of $\mathcal{I}_{\varepsilon}^{GW}$ from Eqs. (\ref{eq:Ie-GW-thetaFn-Infinite-1}) and (\ref{eq:Ie-GW-tanh-Infinite-4}) for the Heaviside step function and the tanh profiles.

\color{black}

\section{A quick look at the integral $\mathcal{I}_{\varepsilon}^{GW}$ for periodic GW memory}\label{Appn:IeGW-periodic-GWM}

For the sake of comparing our results with the one obtained in \cite{Xu:2020pbj} for periodic GW memory, we estimate their $\mathcal{I}_{\varepsilon}^{GW}$. In particular, their gravitational perturbation denoting function was $f(u)=\mathcal{A}\,\cos{(\omega\,u)}$. We shall consider both the Gaussian and eternal switching functions to understand the nature of the concerned quantity $\mathcal{I}_{\varepsilon}^{GW}$. It should also be noted that the expressions of all the other quantities remain the same in this scenario. For example, the expressions of the integrals $\mathcal{I}_{j}$ and $\mathcal{I}_{\varepsilon}^{M}$ for the Gaussian switching are again given by Eqs. (\ref{eq:Ij-Gaussian-3}) and (\ref{eq:Ie-M-Gaussian-2}). On the other hand, $\mathcal{I}_{j}$ and $\mathcal{I}_{\varepsilon}^{M}$ vanish in this scenario with eternal switching.

\subsection{Evaluation of $\mathcal{I}_{\varepsilon}^{GW}$ with the Gaussian switching}\label{Appn:IeGW-cos-Gsn}

We first consider situation with the Gaussian switching function $\kappa(\tau_{j})=e^{-\tau_{j}^2/(2\sigma^2)}$. Considering the GW memory specified by the function $f(u)=\mathcal{A}\,\cos{(\omega\,u)}$ one can obtain the expression of the part of the Wightman function generated purely due to this GW as
\begin{eqnarray}\label{eq:Ie-GW-cos-Gaussian-1}
    G_{W_{GW}}(X_{B},X_{A}) &=& -\frac{\mathcal{A}(\Delta x^2-\Delta y^2)}{4\pi^2}\times\frac{\bigg[\sin{\left\{\omega\,(u_{B}-u_{A})/2\right\}}\,\cos{\left\{\omega\,(u_{B}+u_{A})/2\right\}}\bigg]}{\omega\,\Delta u/2}\times \frac{1}{(\sigma_{M}+i\,\epsilon\,\Delta u)^2}~\nonumber\\
    ~&=& -\frac{\mathcal{A}\,d^2}{4\pi^2}\times\frac{\bigg\{\sin{\left(\frac{\omega\,\Bar{\eta}}{2}\right)}\,\cos{\left(\frac{\omega\,\Bar{\xi}}{2}\right)}\bigg\}}{\omega\,\Bar{\eta}/2}\times \frac{1}{(\Bar{\eta}^2-d^2-i\,\epsilon\,\Bar{\eta})^2}~.
\end{eqnarray}
As discussed earlier from this expression one can get the form of $G_{W_{GW}}(X_{A},X_{B})$ through the relation $G_{W_{GW}}(X_{A},X_{B})=G_{W_{GW}}(X_{B},X_{A})^{*}$. Then the integral $\mathcal{I}_{\varepsilon}^{GW}$ from Eq. (\ref{eq:Ie-general-decomp}) with the Gaussian switching function and with the help of Eq. (\ref{eq:Ie-GW-cos-Gaussian-1}) is represented as
\begin{eqnarray}\label{eq:Ie-GW-cos-Gaussian-2}
    \mathcal{I}_{\varepsilon}^{GW} = \frac{\mathcal{A}\,d^2}{4\pi^2}\int_{0}^{\infty}d\Bar{\eta}\int_{-\infty}^{\infty}d\Bar{\xi}\,e^{-\frac{\Bar{\eta}^2+\Bar{\xi}^2}{4\sigma^2}+i\Delta E\,\Bar{\xi}}\frac{\bigg\{\sin{\left(\frac{\omega\,\Bar{\eta}}{2}\right)}\,\cos{\left(\frac{\omega\,\Bar{\xi}}{2}\right)}\bigg\}}{\omega\,\Bar{\eta}/2}\times \frac{1}{(\Bar{\eta}^2-d^2-i\,\epsilon\,\Bar{\eta})^2}~.
\end{eqnarray}
The integral over the variable $\Bar{\xi}$ can be done using the general integration formulas of Gaussian functions. After carrying out this integration the previous integral will take the form of:
\begin{eqnarray}\label{eq:Ie-GW-cos-Gaussian-3}
    \mathcal{I}_{\varepsilon}^{GW} &=& \frac{\mathcal{A}\,d^2\,\sigma\,\left(e^{2 \sigma^2 \omega \Delta E}+1\right) e^{-\sigma^2 (\omega+2 \Delta E)^2/4}}{2\pi^{3/2}\omega}\int_{0}^{\infty}\frac{d\Bar{\eta}}{\Bar{\eta}}\, \frac{e^{-\Bar{\eta}^2/4\sigma^2}\,\sin{\left(\frac{\omega\,\Bar{\eta}}{2}\right)}}{(\Bar{\eta}^2-d^2-i\,\epsilon\,\Bar{\eta})^2}~
\end{eqnarray}
This integral is doable and one can take the help of numerical methods to obtain the result, which gives outcomes (depicted in Fig. \ref{fig:Ie-Gsn-cos-vDE}) with characteristics similar to the ones obtained for the cases of burst profiles without asymptotic memory, i.e., the Gaussian and sech-squared profiles, provided in Sec. \ref{sec:entanglement-harvesting}. The expression of $\mathcal{I}_{\varepsilon}^{GW}$ as obtained in Eq. (\ref{eq:Ie-GW-cos-Gaussian-3}) signifies the result provided in \cite{Xu:2020pbj}, which also agrees qualitatively with our considered system of burst profiles without memory.

%%%%%%%%%%%%%%%%%%%%%%%%%%%%%%%%%%%%%%%%%%%%%%%%%%%%%%%
\begin{figure*}
\centering
\includegraphics[width=0.47\linewidth]{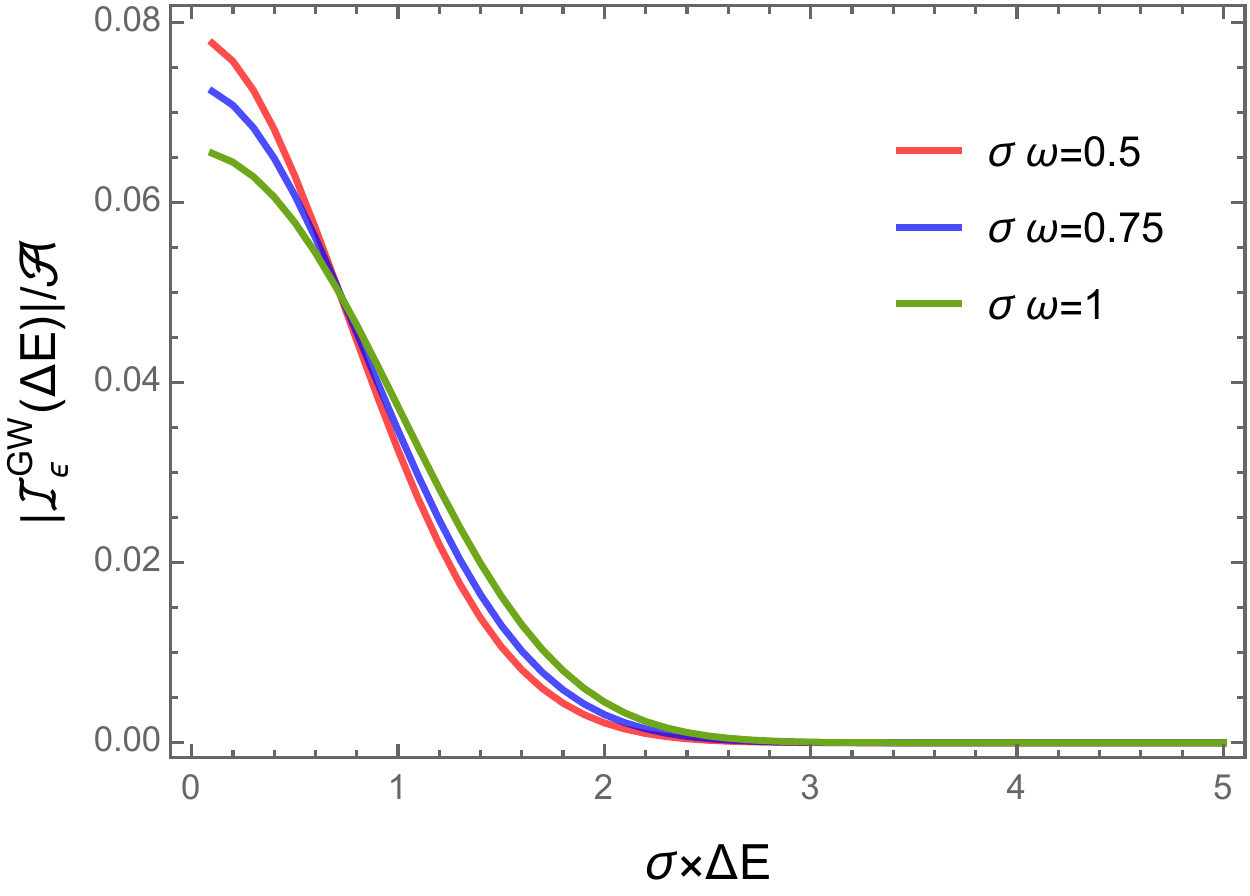}
\hskip 10pt
\includegraphics[width=0.47\linewidth]{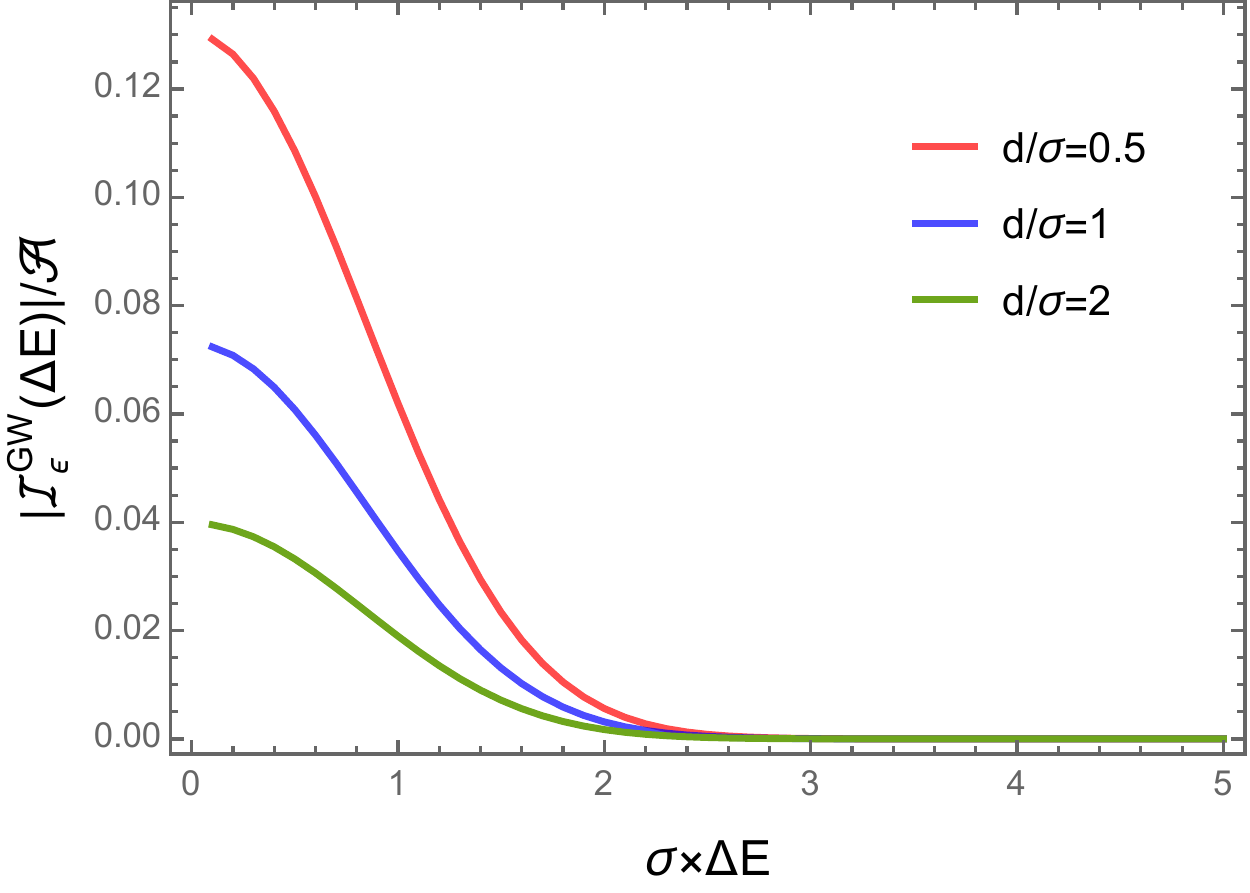}
\caption{The modulus of the quantity $\mathcal{I}_{\varepsilon}^{GW}(\Delta E)$ is plotted as a function of the dimensionless detector transition energy $(\sigma\,\Delta E)$ for $f(u)=\mathcal{A} \cos{(\omega\,u)}$. Both of the above plots correspond to detectors interacting with the background field with Gaussian switching functions $\kappa(\tau_{j})=e^{-\tau_{j}^2/(2\sigma^2)}$. In the left plot, different curves correspond to different $\sigma\,\omega$, while $d/\sigma=1$ is fixed. In the right plot, different curves correspond to different distances $d/\sigma$ between the static detectors, and we have fixed $\sigma\,\omega=0.75$. These plots are presented for a comparison with the ones obtained in our case (Figs. \ref{fig:Ie-Gsn-vDE}, \ref{fig:Ie-Inf-Gsn-sech2-vDE}, and \ref{fig:Ie-Inf-tanh-vDE}).}\label{fig:Ie-Gsn-cos-vDE}
\end{figure*}
%%%%%%%%%%%%%%%%%%%%%%%%%%%%%%%%%%%%%%%%%%%%%%%%%%%%%%%%

\subsection{Evaluation of $\mathcal{I}_{\varepsilon}^{GW}$ with eternal switching}\label{Appn:IeGW-cos-Infty}

Let us now check whether with an eternal switching $\kappa(\tau_{j})=1$, the periodic GW memory $f(u)=\mathcal{A}\,\cos{(\omega\,u)}$ provides similar result. In this scenario, using the Wightman function from Eq. (\ref{eq:Ie-GW-cos-Gaussian-1}), the integral $\mathcal{I}_{\varepsilon}^{GW}$ signifying the non-local entangling term due to the GW memory can be expressed as
\begin{eqnarray}\label{eq:Ie-GW-cos-Infinite-1}
    \mathcal{I}_{\varepsilon}^{GW} = \frac{\mathcal{A}\,d^2}{4\pi^2}\int_{0}^{\infty}d\Bar{\eta}\int_{-\infty}^{\infty}d\Bar{\xi}\,e^{i\Delta E\,\Bar{\xi}}\frac{\bigg\{\sin{\left(\frac{\omega\,\Bar{\eta}}{2}\right)}\,\cos{\left(\frac{\omega\,\Bar{\xi}}{2}\right)}\bigg\}}{\omega\,\Bar{\eta}/2}\times \frac{1}{(\Bar{\eta}^2-d^2-i\,\epsilon\,\Bar{\eta})^2}~.
\end{eqnarray}
We first perform the $\Bar{\xi}$ integration here. It is to be noted that the $\cos{(\omega\,\Bar{\xi}/2)}$ term can be expressed as a sum of exponential terms. Those terms multiplied by $e^{i\Delta E\,\Bar{\xi}}$ will give rise to two Dirac delta distributions with two different arguments after the integration over $\Bar{\xi}$. One of these Dirac deltas will be $\delta(\Delta E+\omega/2)$, which cannot contribute to the result as both $\Delta E$ and $\omega$ are considered to be positive real parameters. Then the remaining term will be
\begin{eqnarray}\label{eq:Ie-GW-cos-Infinite-2}
    \mathcal{I}_{\varepsilon}^{GW} = \frac{\mathcal{A}\,d^2}{2\pi\,\omega}\,\delta(\Delta E-\omega/2)\int_{0}^{\infty}d\Bar{\eta}~\frac{\sin{\left(\frac{\omega\,\Bar{\eta}}{2}\right)}}{\Bar{\eta}}\times \frac{1}{(\Bar{\eta}^2-d^2-i\,\epsilon\,\Bar{\eta})^2}~.
\end{eqnarray}
Here the integral over $\Bar{\eta}$ turns out to be finite. However, there is a Dirac delta distribution $\delta(\Delta E-\omega/2)$ sitting outside. This quantity signifies that the integral $\mathcal{I}_{\varepsilon}^{GW}$ will vanish, so will the concurrence, whenever $\Delta E\neq \omega/2$, and it will be infinite when $\Delta E=\omega/2$. Thus the concurrence exists only when $\Delta E=\omega/2$. This is unlike our GW burst scenarios.\vspace{0.4cm}
\end{widetext}

%\-lobibliographystyle{apsrev}
\bibliographystyle{utphys1.bst}

\bibliography{bibtexfile}

\end{document}